\documentclass[journal]{IEEEtran}
\ifCLASSINFOpdf

\else
\fi
\usepackage{algorithm}
\usepackage{algorithmic}
\usepackage{booktabs}
\usepackage{makecell}
\usepackage{multirow}
\usepackage{multicol}
\usepackage{tikz}
\usepackage{caption}
\definecolor{bblue}{HTML}{3371FF}
\definecolor{ggreen}{HTML}{9CFF33}
\definecolor{ppurple}{HTML}{4C0B5F}
\definecolor{ppink}{HTML}{FF0080}
\definecolor{rrq5}{HTML}{6092B9}
\usepackage{pgfplots}
\usepackage{cite}
\usepackage{xcolor}
\usepackage{amsfonts}
\usepackage{amsmath}
\usepackage{amsthm}
\usepackage{mathtools}
\usepackage{commath}
\newtheorem{rmk}{Remark}

\usepackage{graphicx}
\usepackage{amssymb}
\usepackage{subcaption}

\usepackage[normalem]{ulem}
\usepackage[hidelinks]{hyperref}
\usepackage[nameinlink, capitalise, noabbrev]{cleveref}
\usepackage{xcolor,cite,etoolbox}
\makeatletter 
\pretocmd\@bibitem{\color{black}\csname keycolor#1\endcsname}{}{\fail}
\newcommand\citecolor[1]{\@namedef{keycolor#1}{\color{black}}}
\makeatother
\citecolor{paudel2019loss}
\citecolor{fiftyshades2016}
\citecolor{fewer17}
\citecolor{pan2008}
\citecolor{Tran2021}
\citecolor{wang2017signed}
\citecolor{kim2018side}
\citecolor{derr2018signed}
\citecolor{huang2019signed}
\citecolor{li2020learning}
\citecolor{liu2021signed}
\citecolor{tang2015line}
\citecolor{mikolovskipgram}
\citecolor{chen2020neural_HGCN}
\citecolor{yuan2017sne}
\citecolor{leskovec2010predicting}

\begin{document}
\usepgfplotslibrary{groupplots}
%
\title{\textsf{SiReN}: Sign-Aware Recommendation\\Using Graph Neural Networks}
%
%
%

\author{Changwon Seo, Kyeong-Joong Jeong, Sungsu Lim, \IEEEmembership{Member, IEEE}, and Won-Yong Shin, \IEEEmembership{Senior Member, IEEE}
\thanks{ C. Seo, K.-J. Jeong, and W.-Y. Shin are with the School of Mathematics and Computing (Computational Science and Engineering), Yonsei University, Seoul 03722, Republic of Korea (e-mail: \{changwoni, jeongkj, wy.shin\}@yonsei.ac.kr).}
\thanks {S. Lim is with the Department of Computer Science and Engineering, Chungnam National University, Daejeon 34134, Republic of Korea (e-mail: sungsu@cnu.ac.kr).}
\thanks{\textit{(Corresponding author: Won-Yong Shin.)}}}
\maketitle

\begin{abstract}
In recent years, many recommender systems using network embedding (NE) such as graph neural networks (GNNs) have been extensively studied in the sense of improving recommendation accuracy. However, such attempts have focused mostly on utilizing only the information of positive user--item interactions with high ratings. Thus, there is a challenge on how to make use of {\em low rating} scores for representing users' preferences since low ratings can be still informative in designing NE-based recommender systems. In this study, we present \textsf{SiReN}, a new {\em \underline{Si}gn-aware} \underline{Re}commender system based on GN\underline{N} models. Specifically, \textsf{SiReN} has three key components: 1) constructing a {\em signed} bipartite graph for more precisely representing users' preferences, which is split into two edge-disjoint graphs with {\em positive} and {\em negative} edges each, 2) generating two embeddings for the partitioned graphs with positive and negative edges via a GNN model and a multi-layer perceptron (MLP), respectively, and then using an attention model to obtain the final embeddings, and 3) establishing a sign-aware Bayesian personalized ranking (BPR) loss function in the process of optimization. Through comprehensive experiments, we empirically demonstrate that \textsf{SiReN} consistently outperforms state-of-the-art NE-aided recommendation methods.
\end{abstract}


%
\IEEEpeerreviewmaketitle

\section{Introduction}
%
%
%
%

 


\subsection{Background and Motivation}
\IEEEPARstart{R}{ecommender} systems have been widely advocated as a way of providing suitable recommendation solutions to customers in various fields such as e-commerce, advertising, and social media sites. One of the most important and popular techniques in recommender systems is collaborative filtering (CF), which computes similarities between users and items from historical interactions (e.g., clicks and purchases) to suggest relevant items to users by assuming that users who have behaved similarly with each other exhibit similar preferences for items \cite{hsieh2017collaborative,ebesu2018collaborative,han2019adaptive}. Moreover, following up the great success of network embedding (NE), also known as network representation learning, considerable research attention has been paid to NE-based recommender systems that attempt to model high-order connectivity information from user--item interactions viewed as a bipartite graph \cite{gori2007itemrank, yang2018hop, gao2018bine,chen2019collaborative}.\footnote{In the following, we use the terms ``graph" and ``network" interchangeably.} In recent years, {\em graph neural networks (GNNs)} \cite{kipf2017semi,hamilton2017inductive, xu2018powerful,velivckovic2017graph ,wu2019survey} have emerged as a powerful neural architecture to learn vector representations of nodes and graphs. By virtue of great prowess in solving various downstream machine learning problems, GNN-based recommender systems \cite{ying2018graph,zheng2018spectral,wang2019neural,wang2020multi,wu2020joint,chen2020revisiting,he2020lightgcn} have also been developed for improving the recommendation accuracy.

\begin{figure}[t]
	\centering
 	\includegraphics[width=\columnwidth]{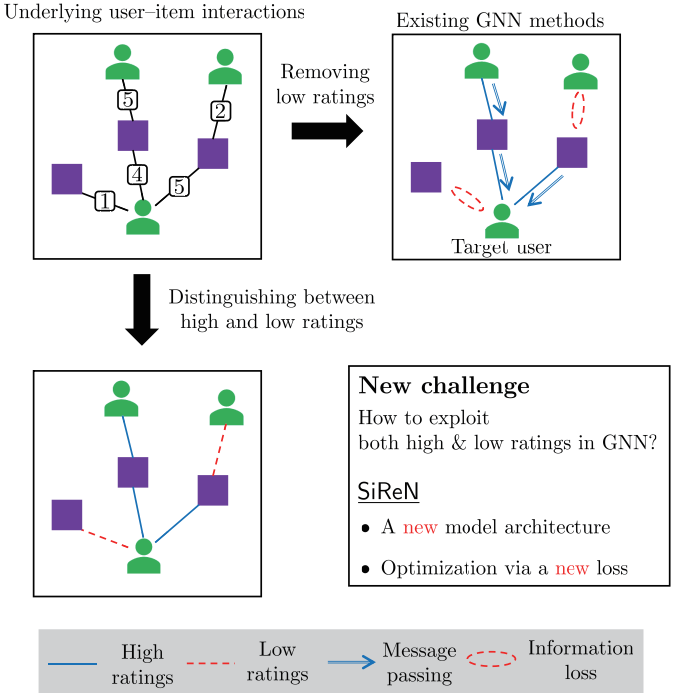}
	\caption{New challenges in GNN-based recommender systems.}
	\label{fig:intro}
\end{figure}

GNN models are basically trained by aggregating the information of direct neighbor nodes via message passing under the {\em homophily} (or {\em assortativity}) assumption that a target node and its neighbors are similar to each other \cite{wu2019survey}.  Due to the neighborhood aggregation mechanism, existing literature posits that high homophily of the underlying graph is a necessity for GNNs to achieve good performance especially on node classification \cite{zhu2020beyond,pei2020geom,zhu2020graph}. On the other hand, in recommender systems, while users' feedback on many online websites (e.g., {\em likes/dislikes} on YouTube and high/low ratings on Amazon) can be positive and negative, existing GNN-based recommender systems overlook the existence of negative feedback \color{black}(e.g., {\em dislikes} on YouTube and low ratings on Amazon) \color{black}due to their ease of modeling. Precisely, most GNN-based approaches utilize only positive feedback by removing the negative interactions in order to exhibit strong homophily in neighbors (see Fig. \ref{fig:intro}).\footnote{A graph is usually referred to as homophilous (or assortative) if connected nodes are much more likely to have the same class label than if edges are independent of labels. In our study, a {\em homophilous bipartite} graph assumes that user--item relations are significantly more likely to be positive.} It is worthwhile to note that, despite the remarkable performance boost by existing GNN-based recommender systems, the low ratings can be still informative. This is because such information expresses signs of what users {\em dislike}. In other words, full exploitation of two types of feedback in GNNs may have the potential to further improve the recommendation performance, which remains a new design challenge.

\subsection{Main Contributions}

\color{black}
CF in recommender systems has been often studied by regarding low ratings as {\em implicit} negative feedback (see \cite{paudel2019loss, fiftyshades2016, fewer17,Tran2021}). However, such interpretations may not be straightforwardly extended to GNN-based recommender systems due to their inherent architecture including neighborhood aggregation. In this context, \color{black}even with the wide applications of GNNs to recommender systems \cite{ying2018graph,zheng2018spectral,wang2019neural,wang2020multi,wu2020joint,chen2020revisiting,he2020lightgcn}, a question that arises naturally is: ``How can we make use of \color{black}implicit \color{black} negative feedback (i.e., low rating scores) for representing users' preferences via GNNs?". In this paper, to answer this question, we introduce \textsf{SiReN}, a {\em \underline{Si}gn-aware} \underline{Re}commender system based on GN\underline{N} models. To this end, \textcolor{black}{as illustrated in Fig.~\ref{fig:intro}}, we design and optimize our new GNN-aided learning model while distinguishing users' positive and negative feedback.

Our proposed \textsf{SiReN} method includes three key components: 1) signed graph construction and partitioning, 2) model architecture design, and 3) model optimization. First, to overcome the primary problem of existing GNN-based recommender systems \cite{wang2019neural, chen2020revisiting, he2020lightgcn} that fail to learn both positive and negative relations between users and items, we start by constructing a {\em signed} bipartite graph, which is split into two edge-disjoint graphs with {\em positive} and {\em negative} edges each. This signed graph construction and partitioning process enables us to more distinctly identify users' preferences for observed items. Second, we show how to design our model architecture for discovering embeddings of all users and items in the signed bipartite graph. Although several GNN models were introduced in \cite{derr2018signed, huang2019signed} for signed unipartite graphs, simply applying them to recommender systems, corresponding to user--item bipartite graphs, may not be desirable. This is because such GNN models were built upon the assumption of balance theory \cite{heider1946attitudes}, which implies that ``{\em the friend of my friend is my friend}" and ``{\em the enemy of my enemy is my friend}". However, the balance theory no longer holds in recommender systems since users' preferences cannot be dichotomous.  In other words, users are likely to have dissimilar preferences even if they dislike the same item(s). This motivates us to design our own model architecture for each partitioned graph, rather than employing existing GNN approaches based on the balance theory. Concretely, \textsf{SiReN} contains three learning models. For the graph with positive edges, we employ a GNN model suit for recommender systems. For the graph with negative edges, we adopt a multi-layer perceptron (MLP) due to the fact that negative edges can weaken the homophily and thus message passing to such dissimilar nodes would not be feasible. To obtain the final embeddings, we then use an attention model that learns the importance of two embeddings generated from GNN and MLP models. Third, as our objective function in the process of optimization, we present a {\em sign-aware} Bayesian personalized ranking (BPR) loss function, which is built upon the original BPR loss \cite{BPRMF09} widely used in recommender systems. More specifically, unlike the original BPR loss, our objective function takes into account two types of observed items, including both positive and negative relations between users and items, as well as unobserved ones.

To validate the superiority of our \textsf{SiReN} method, we comprehensively perform empirical evaluations using various real-world datasets. Experimental results show that our method consistently outperforms state-of-the-art GNN methods for top-$K$ recommendation in terms of several recommendation accuracy metrics. Such a gain is possible owing to the use of low rating information along with judicious model design and optimization. We also empirically validate the effectiveness of MLP in comparison with other model architectures used for the graph with negative edges. Additionally, our experimental results demonstrate the robustness of our \textsf{SiReN} method to more challenging interaction sparsity levels.

It is worth noting that our method is GNN-{\em model-agnostic} and thus any competitive GNN architectures can be appropriately chosen for potentially better performance. The main technical contributions of this paper are four-fold and summarized as follows:
\begin{itemize}
    \item We propose \textsf{SiReN}, a novel GNN-aided recommender system that makes full use of the user--item interaction information after signed graph construction and partitioning;
    \item We design our model architecture using three learning models in the sense of utilizing sign awareness so as to generate embeddings of users and items;
    \item We establish the sign-aware BPR loss as our objective function;
    \item We validate \textsf{SiReN} through extensive experiments using three real-world datasets while showing the superiority of our method over state-of-the-art NE-aided methods under diverse conditions.
\end{itemize}

\subsection{Organization and Notation}
The remainder of this paper is organized as follows. In Section II, we present prior studies related to our work. In Section III, we explain the methodology of our study, including basic settings and an overview of our \textsf{SiReN} method. Section IV describes technical details of the proposed method. Comprehensive experimental results are shown in Section V. Finally, we provide a summary and concluding remarks in Section VI.

\section{Related work}
The method that we proposed in this study is related to \color{black}three \color{black} broader research lines, namely standard CF approaches\color{black}, \color{black} NE-based recommendation approaches\color{black}, and NE approaches for signed graphs.\color{black}

\subsection{Standard CF Approaches}
As one of the most popular techniques, CF in recommender systems aims to capture the relationships between users and items from historical interactions (e.g., ratings and purchases) by discovering learnable vector representations for users and items based on a rating matrix \cite{koren2008factorization,ebesu2018collaborative}. Matrix factorization (MF) \cite{koren2009matrix} decomposes the user--item interaction matrix into a product of two low-dimensional matrices and models their similarities with the dot product of two matrices. ANLF \cite{luo2015nonnegative} was designed based on a non-negative MF (NMF) model \cite{lin2007projected} using the alternating direction method. In DMF \cite{xue2017deep} and NCF \cite{he2017neural}, MF models with neural network architectures were proposed to project users and items into a latent structured space. Moreover, SLIM \cite{ning2011slim} proposed a sparse linear model by directly reconstructing the low-density user--item interaction matrix. To capture complex relationships between items in sparse datasets, FISM \cite{kabbur2013fism} and NPE \cite{nguyen2018npe} showed how to learn item--item similarities as the product of two latent factor matrices. While MF-based models such as \cite{xue2017deep, he2017neural} rely on the dot product of user and item latent vectors as a similarity measure, CML \cite{hsieh2017collaborative} showed how to learn a joint metric space to encode not only users' preferences but also user--user and item--item similarities in the Euclidean distance. In NAIS \cite{he2018nais}, the attention mechanism was incorporated to obtain the importance of each item from user--item interactions for preference prediction. DLMF \cite{deng2016deep} designed a trust-aware recommender system that leverages deep learning to determine the initialization in MF by synthesizing the users' interests and their trusted friends' interests in addition to the user--item interactions.

\subsection{NE-Based Recommendation Approaches}
Recently, it has been comprehensively studied how to develop recommender systems using NE. While standard CF approaches are capable of only modeling the first-order connectivity between users and items, NE-based approaches aim to exploit {\em high-order proximity} among users and items through the user--item bipartite graph structure \cite{wang2019neural, he2020lightgcn}. BiNE \cite{gao2018bine} and CSE \cite{chen2019collaborative} were developed based on random walks in order to infer similar users (or items) in the underlying bipartite graph (i.e., user--item interactions). 

Moreover, encouraged by the success of GNNs in solving many graph mining tasks \cite{wu2019survey}, GNN-based recommender systems have more recently emerged as promising techniques \cite{van2018graph, ying2018graph, zheng2018spectral, chen2020revisiting, zhang2019inductive, wang2019neural, he2020lightgcn, wang2020multi, wu2020joint }. Existing GNN models for top-$K$ recommendation were generally developed using implicit feedback, which treats observed user--item interactions as positive relations. GC-MC \cite{van2018graph} presented a graph autoencoder including graph convolutions for rating matrix completion. PinSage \cite{ying2018graph} showed a scalable GNN framework developed in production at Pinterest by improving upon GraphSAGE \cite{hamilton2017inductive} along with several aggregation functions. Spectral CF \cite{zheng2018spectral} presented a spectral convolution operation to learn the rich information of connectivity between users and items in the spectral domain. As state-of-the-art NE-based recommender systems using only user--item interactions as input, NGCF \cite{wang2019neural}, LR-GCCF \cite{chen2020revisiting}, and LightGCN \cite{he2020lightgcn} were developed based on GCN \cite{kipf2017semi}, the first attempt to apply convolutions to graph domains, while performing layer aggregation to solve the oversmoothing problem. IGMC \cite{zhang2019inductive} proposed an inductive GNN-based matrix completion model that learns local subgraphs in the underlying user--item interaction matrix. To explore users' latent purchasing motivations (e.g., cost effectiveness and appearance), MCCF \cite{wang2020multi} presented a CF approach based on GNNs for generating multiple user/item embeddings and then combining them via attention mechanisms. AGCN \cite{wu2020joint} proposed a GCN approach for joint item recommendation and attribute inference in an attributed user--item bipartite graph with incomplete attribute values. 

\color{black}In addition to various GNN architectures in \cite{van2018graph, ying2018graph, zheng2018spectral, chen2020revisiting, zhang2019inductive, wang2019neural, he2020lightgcn, wang2020multi, wu2020joint } for recommender systems, to \color{black} alleviate the sparsity of user--item interactions, each user's local neighbors' preferences in social networks were also utilized in \cite{wu2019neural, fan2019graph, wu2020diffnet++} for better user embedding modeling, thus enabling us to improve the recommendation accuracy. \color{black}An approach for formulating feature-aware recommendation from the review information via a signed hypergraph convolutional network was also presented in \cite{chen2020neural_HGCN}.

\subsection{NE Approaches for Signed Networks}
The design of NE in {\em signed} networks has garnered considerable attention while primarily solving the problem of link sign prediction \cite{wang2017signed, kim2018side,derr2018signed,huang2019signed,li2020learning,liu2021signed}, which is to predict unobserved signs of existing edges \cite{leskovec2010predicting}. Specifically, NE methods in \cite{yuan2017sne}, \cite{kim2018side} for signed networks were developed by formulating their own likelihood functions based on generated random walk sequences. Moreover, GNN-aided NE approaches for signed networks were presented in \cite{derr2018signed,huang2019signed,li2020learning} while being guided by the structural balance theory \cite{heider1946attitudes}. SGCN \cite{derr2018signed} generalized GCN \cite{kipf2017semi} to signed networks by generating both balanced and unbalanced embeddings. SiGAT \cite{huang2019signed} was built upon GAT \cite{velivckovic2017graph} along with motif-based GNNs. SNEA \cite{li2020learning} presented a universal way of leveraging the graph attention mechanism to aggregate more important information through both positive and negative edges. By adopting the k-group theory beyond the balance theory, GS-GNN \cite{liu2021signed} designed a dual GNN architecture to learn global and local representations.

\color{black}

\subsection{Discussion}
Despite the aforementioned contributions, leveraging {\em explicit feedback} data (i.e., user--item interactions with ratings) in NE-based approaches has been largely underexplored in the literature. To learn vector representations for users and items based on the NE, it is common to utilize only positive user--item interactions as observed data when explicit feedback data are given (see \cite{chen2019collaborative,wu2019neural,wu2020diffnet++} and references therein). However, negative interactions with low ratings can be still informative since such information shows signs of what users {\em dislike}. It remains open how to make use of low ratings in designing NE-aided recommender systems.

\color{black}
On the other hand, as aforementioned, several GNN models \cite{derr2018signed,huang2019signed,li2020learning,liu2021signed} have been actively developed to learn vector representations of nodes in {\em signed} (unipartite) graphs with both positive and negative edges based on the balance theory or its variant. However, it is worth noting that these models focus primarily on such networks that exhibit friend/foe (or trust/distrust) relationships. This implies that they do not always hold in recommender systems due to the fact that users' preferences cannot be dichotomous, i.e., a behavior of users disliking similar items does not alway imply the same degree of user preferences about items. Therefore, adopting such GNN methods designed for signed graphs would not be desirable to capture different levels of user preferences in recommender systems. \color{black}

\begin{figure*}[t]
	\centering
	\includegraphics[width=1.8\columnwidth]{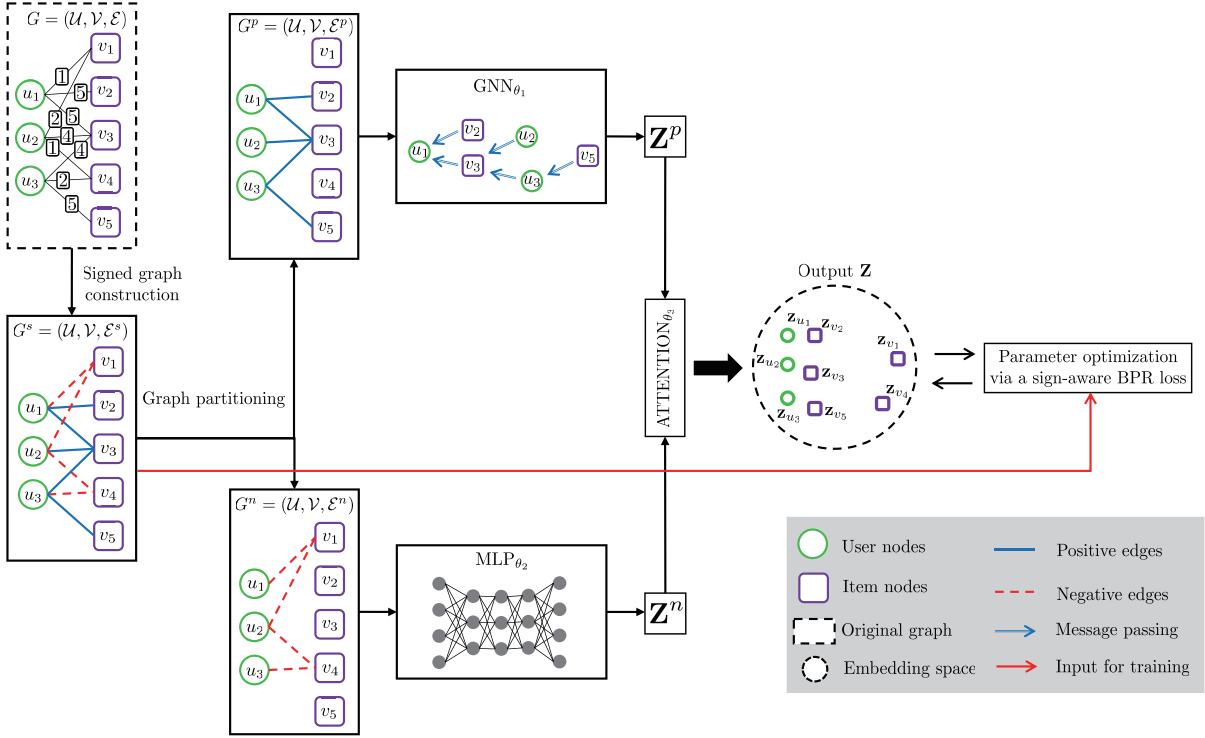}
	\caption{The schematic overview of our \textsf{SiReN} method.}
	\label{fig:overview}
\end{figure*}

\section{Methodology}
In this section, we describe our network model with basic settings. Next, we explain an overview of the proposed \textsf{SiReN} method as a solution to the problem of making full use of low ratings in GNN models. 
\subsection{Network Model and Basic Settings}
In recommender systems, the basic input is the historical user--item interactions with ratings, which is represented as a weighted bipartite graph. Let us denote the underlying bipartite graph as $G =( \mathcal{U}, \mathcal{V}, \mathcal{E})$, where $\mathcal{U}$ and $\mathcal{V}$ are the set of $M$ users and the set of $N$ items, respectively, and $\mathcal{E}$ is the set of weighted edges between $\mathcal{U}$ and $\mathcal{V}$. A weighted edge $(u,v,w_{uv})\in \mathcal{E}$ can be interpreted as the rating $w_{uv}$ with which the user $u\in \mathcal{U}$ has given to the item $v\in \mathcal{V}$. We assume $G$ to be a static network without repeated edges, where ratings (i.e., user preferences) do not change over time. 

In our study, we aim at designing a new GNN-aided recommender system for improving the accuracy of top-$K$ recommendation by making full use of the user--item rating information in $G$, including {\em low ratings} that have not been explored by conventional GNN-based recommender systems \cite{he2020lightgcn,chen2020revisiting}, without any side information.

\subsection{Overview of \textsf{SiReN}}
In this subsection, we explain our methodology along with the overview of the proposed \textsf{SiReN} method. We recall that our study is motivated by the fact that recent recommender systems built upon GNN models such as \cite{he2020lightgcn,chen2020revisiting} take advantage of only high rating scores as observed data by deleting some edges, corresponding to low ratings (e.g., the rating scores of 1 and 2 in the 1--5 rating scale), in the set $\mathcal{E}$ over the weighted bipartite graph $G$ \cite{chen2019collaborative, wu2019neural,wu2020diffnet++}. This is because such a removal of low ratings from $G$ enables us to aggregate the positively connected neighbors via message passing in GNNs. However, the set of negative interactions indicates what users {\em dislike} and thus is still quite informative. In other words, it remains open how to fully exploit the rating information in building GNN-based recommender systems as recently developed models fail to capture the effect of low rating scores. 

To tackle this challenge, we present \textsf{SiReN}, a new {\em sign-aware} recommender system using GNNs, which is basically composed of the following three core components (refer to Fig. \ref{fig:overview}): 
\begin{itemize}
\item signed bipartite graph construction and partitioning 
\item embedding generation for each partitioned graph
\item optimization via a sign-aware BPR loss.
\end{itemize}

First, we describe how to construct a signed bipartite graph $G^s$ that enables us to more distinctly identify users' preferences based on all the user--item interactions. More specifically, we construct $G^s=(\mathcal{U}, \mathcal{V}, \mathcal{E}^s)$ with a parameter $w_o>0$, representing a criterion for dividing high and low ratings, where
\begin{align}
\label{align_signededges}
\mathcal{E}^s= \big\{ (u,v,w_{uv}^{s})\big|w_{uv}^{s}=w_{uv}-w_{o},~(u,v,w_{uv})\in \mathcal{E} \big\}.
\end{align}
Here, $w_o$ can be determined according to characteristics (e.g., the rating scale and the popularity distribution of items) of a given dataset; from an algorithm design perspective, we assume that a user $u$ likes an item $v$ if $w_{uv}^s>0$ and he/she dislikes $v$ otherwise, where $w_{uv}^s=w_{uv}-w_o$ corresponds to the edge weight in the signed bipartite graph $G^s$. \color{black} Note that this signed graph construction is not necessary if we deal with originally signed bipartite graphs (e.g., {\em like/dislike} rating systems on YouTube). 
\color{black}Basically, GNN models are trained by aggregating the information of neighbor nodes under the {\em homophily} assumption \cite{wu2019survey}. However, due to the fact that the signed graph $G^s$ includes negative edges (i.e., interactions with low ratings), aggregating the information of such negatively connected neighbors may not be desirable. As illustrated in Fig. \ref{fig:overview}, to more delicately capture each relation of positively and negatively connected neighbors, we then {\em partition} the signed bipartite graph $G^s$ into two edge-disjoint graphs $G^p=(\mathcal{U}, \mathcal{V}, \mathcal{E}^p)$ and $G^n=(\mathcal{U}, \mathcal{V}, \mathcal{E}^n)$, consisting of the set of positive edges and negative edges, respectively. Here, it follows that $\mathcal{E}^s=\mathcal{E}^p\cup \mathcal{E}^n$, where
\begin{subequations}
\begin{align}
\label{graph_split}
\mathcal{E}^p&= \big\{ (u,v,w_{uv}^{s})\big|~w_{uv}^s>0,~(u, v, w_{uv}^{s})\in \mathcal{E}^s \big\}\\
\mathcal{E}^n&= \big\{ (u,v,w_{uv}^{s})\big|~w_{uv}^s<0,~(u, v, w_{uv}^{s})\in \mathcal{E}^s \big\}.
\end{align}
\end{subequations}
The purpose of this graph partitioning is to make the graphs $G^p$ and $G^n$, respectively, assortative and disassortative so that each partitioned graph is used as input to the most appropriate learning model.

Second, we describe how to generate embeddings of $M$ users and $N$ items along with three learning models in our \textsf{SiReN} method. Using the graph $G^p$ having positive edges, we adopt a GNN model suit for recommender systems (e.g., \cite{he2020lightgcn,chen2020revisiting}) to calculate embedding vectors ${\bf Z}^p\in \mathbb{R}^{(M+N)\times d}$ for the nodes in $\mathcal{U}\cup \mathcal{V}$:
\begin{align}
\label{align_GNN_pos}
{\bf Z}^p=\text{GNN}_{\theta_1}(G^p),
\end{align}
where $d$ is the dimension of the embedding space and $\theta_1$ is the learned model parameters of GNN. On the other hand, we adopt an MLP for the graph $G^n$ to calculate embedding vectors ${\bf Z}^n\in \mathbb{R}^{(M+N)\times d}$ for the same nodes as those in (\ref{align_GNN_pos}):
\begin{align}
\label{align_MLP_neg}
{\bf Z}^n=\text{MLP}_{\theta_2}(G^n),
\end{align}
where $\theta_2$ is the learned model parameters of MLP. We would like to state the following two remarks to explain the model selection according to types of graphs.

\begin{rmk}
\label{rmk_1}
It is worth noting that existing GNN models, built upon message passing architectures, work on the basic assumption of homophily \cite{zhu2020graph}. We recall that users who dislike similar items may not have similar preferences (i.e., the tendency in rating items) with each other. Negative edges in $G^n$ can undermine the effect of homophily and thus message passing to such dissimilar nodes would not be feasible. For these reasons, adopting GNNs in the graph with negative edges may not be desirable, based on the fact that many GNNs fail to generalize to disassortative graphs, i.e., graphs with low levels of homophily \cite{pei2020geom,zhu2020beyond,jin2021node}.
\end{rmk}

\begin{rmk}
\label{rmk_2}
Additionally, note that the MLP architecture itself does not exploit the topological information. However, it does not imply that the connectivity information in the graph $G^n$ is not used at all. In the optimization step, we update the embedding vectors in (\ref{align_GNN_pos}) and (\ref{align_MLP_neg}) by fully taking advantage of the set $\mathcal{E}^n$ in $G^n$ as well as the set $\mathcal{E}^p$ in $G^p$, which will be specified later.
\end{rmk}

Next, let us mention another training model in \textsf{SiReN}, the so-called {\em attention} model. To get the importance of two embeddings ${\bf Z}^p$ and ${\bf Z}^n$, we use the attention mechanism \cite{vaswani2017attention} that learns the corresponding importance $({\bf \alpha}_p,{\bf \alpha}_n)$ as follows:
\begin{align}
\label{align_attn}
({\bf \alpha}^p,{\bf \alpha}^n)=\text{ATTENTION}_{\theta_3} ({\bf Z}^p,{\bf Z}^n),
\end{align}
which results in the final embeddings:
\begin{align}
\label{attn_final_emb}
{\bf Z}=({\bf \alpha}^p{\bf 1}_{\text{attn}}) \odot {\bf Z}^p + ({\bf \alpha}^n{\bf 1}_{\text{attn}}) \odot {\bf Z}^n,	
\end{align}
where ${\bf \alpha}^p,{\bf \alpha}^n\in \mathbb{R}^{(M+N)\times 1}$; ${\bf 1}_{\text{attn}}\in\mathbb{R}^{1\times d}$ is the all-ones vector; $\odot$ denotes the Hadamard (element-wise) product; $\theta_3$ is the learned model parameters of ATTENTION; and each row of ${\bf Z}\in\mathbb{R}^{(M+N)\times d}$ indicates the embedding vector of each node in $\mathcal{U}\cup\mathcal{V}$. 

Third, we turn to the optimization of model parameters $\{\theta_1,\theta_2,\theta_3\}$, which updates the embeddings ${\bf Z}$ accordingly. In our study, we adopt the {\em BPR loss} \cite{BPRMF09}, which has been widely used in recommender systems to comprehensively learn what users prefer from the historical user--item interactions. Nevertheless, simply applying the existing BPR loss to our setting does not precisely capture the relations of negatively connected neighbors; thus, we establish a {\em sign-aware} BPR loss, which is a new BPR-based loss that takes into account both positive and negative relations in the signed bipartite graph $G^s$, while accommodating the sign of edges in $G^s$ as an indicator of what users like and dislike.

In the next section, we shall describe implementation details of the proposed \textsf{SiReN} method.

\begin{algorithm}[t]
\caption{: \textsf{SiReN}}
\label{algorithm_main}
\begin{algorithmic}[1]
 \renewcommand{\algorithmicrequire}{\textbf{Input:}}
 \renewcommand{\algorithmicensure}{\textbf{Output:}}
 \REQUIRE $G$, $w_o$, $\Theta \triangleq \{\theta_1, \theta_2, \theta_3\}$, $N_{\text{neg}}$ (number of negative samples), $\lambda_{\text{reg}}$ (regularization coefficient)
 \ENSURE  $\mathbf{Z}$
 \STATE \textbf{Initialization: }$\Theta \leftarrow $ random initialization
 \STATE /* Graph partitioning */
 \STATE Construction of $G^s$ from $G$ along with $w_o$
 \STATE Splitting $G^s$ into $G^p$ and $G^n$

 \WHILE{not converged}
 \STATE Create \color{black}a training set \color{black} $\mathcal{D}_S$ from $G^s$ with $N_{\text{neg}}$ negative samples
 \FOR {each \color{black}mini-batch \color{black} $\mathcal{D}_S' \subset \mathcal{D}_S$}
 \STATE /* Generation of embeddings */
 \STATE ${\bf Z}^p\leftarrow \text{GNN}_{\theta_1}(G^p)$
 \STATE ${\bf Z}^n\leftarrow \text{MLP}_{\theta_2}(G^n)$
 \STATE $({\bf \alpha}^p,{\bf \alpha}^n)\leftarrow \text{ATTENTION}_{\theta_3} ({\bf Z}^p,{\bf Z}^n)$
 \STATE ${\bf Z}\leftarrow ({\bf \alpha}^p{\bf 1}_{\text{attn}})\odot {\bf Z}^p + ({\bf \alpha}^n{\bf 1}_{\text{attn}})\odot{\bf Z}^n$
 \STATE /* Optimization */
 \STATE $\mathcal{L}_0 \leftarrow \text{\textit{sign-aware} BPR loss}$
 \STATE $\mathcal{L} \leftarrow {\mathcal{L}}_0+\lambda_{\text{reg}}\norm{\bf \Theta}^2$ 
 \STATE Update $\Theta$ by taking one step of gradient descent
 \ENDFOR
 \ENDWHILE
 \RETURN ${\bf Z}$
\end{algorithmic}
\end{algorithm}

\section{Proposed \textsf{SiReN} Method}
In this section, we elaborate on our \textsf{SiReN} method, designed for top-$K$ recommendation. \textsf{SiReN} has the following three key components: 1) constructing a signed bipartite graph $G^s$ for more precisely representing users' preferences, which is split into two graphs $G^p$ and $G^n$ with {\em positive} and {\em negative} edges each, 2) generating two embeddings ${\bf Z}^p$ and ${\bf Z}^n$ for the partitioned graphs with positive and negative edges via a GNN model and an MLP, respectively, and then using the attention model to learn the importance of ${\bf Z}^p$ and ${\bf Z}^n$, and 3) establishing a sign-aware BPR loss function in the process of optimization. The overall procedure of the proposed \textsf{SiReN} method is summarized in Algorithm \ref{algorithm_main}.

As one of main contributions to the design of our method, we start by constructing the signed bipartite graph $G^s$ and then partitioning $G^s$ into two edge-disjoint graphs $G^p$ and $G^n$ for exploiting the relation of positively and negatively connected neighbors.

In the following subsections, we explain how we discover embeddings of all nodes (i.e., users and items) and optimize our model via the sign-aware BPR loss during the training phase.

\subsection{Network Architecture}
In this subsection, we describe how to generate the embeddings of $M$ users and $N$ items in our \textsf{SiReN} method. As stated in Section III-B, \textsf{SiReN} basically contains three learning models, i.e., $\text{GNN}_{\theta_1}$, $\text{MLP}_{\theta_2}$, and $\text{ATTENTION}_{\theta_3}$. To generally indicate either users or items interchangeably, we denote a node in the graphs $G^p$ and $G^n$, which can be in either $\mathcal{U}$ or $\mathcal{V}$, by $x$.

\begin{figure}[t]
	\centering
 	\includegraphics[width=0.7\columnwidth]{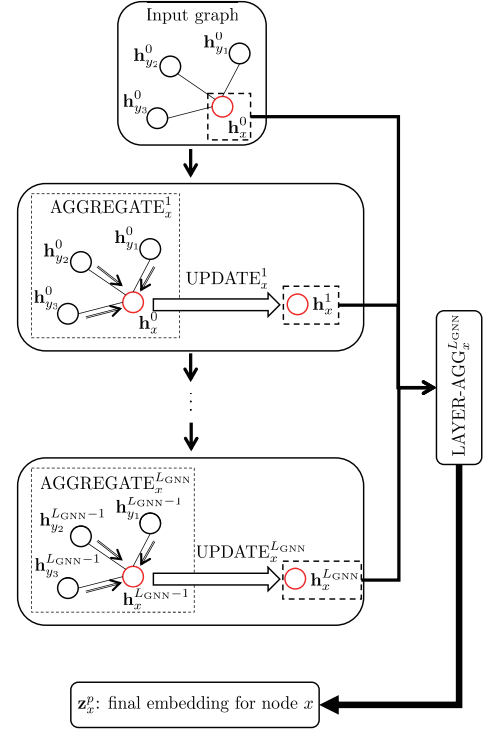}
	\caption{The GNN architecture in our \textsf{SiReN} method, composed of three functions in (\ref{aggregate_})--(\ref{layer_agg}). }
	\label{fig:GNN_}
\end{figure}

First, we describe the GNN model, which is designed to be {\em model-agnostic}. To this end, we show a general form of the message passing mechanism \cite{hamilton2017inductive, gilmer2017neural, xu2018powerful} in which we iteratively update the representation of each node by aggregating representations of its neighbors using two functions, namely $\text{AGGREGATE}_x^{\ell}$ and $\text{UPDATE}_x^{\ell}$, along with model parameters of $\text{GNN}_{\theta_1}$ in (\ref{align_GNN_pos}). Formally, at the $\ell$-th layer of a GNN, $\text{AGGREGATE}^{\ell}_{x}$ aggregates (latent) feature information from the local neighborhood of node $x$ in $G^p$ at the $(\ell-1)$-th GNN layer as follows:
\begin{align}
\label{aggregate_}
{\bf m}_{x}^{\ell} \leftarrow \text{AGGREGATE}^{\ell}_{x}\big(\big\{ {\bf h}_{y}^{\ell-1}\big|y \in \mathcal{N}_{x} \cup \{x\}\big\}\big),	
\end{align}
where ${\bf h}_{x}^{\ell-1}\in\mathbb{R}^{1\times d_{\text{GNN}}^{\ell-1}}$ denotes the $d_\text{GNN}^{\ell-1}$-dimensional latent representation vector of node $x$ at the $(\ell-1)$-th GNN layer, $\mathcal{N}_{x}$ is the set of neighbor nodes of $x$ in $G^p$, and ${\bf m}_{x}^{\ell}\in\mathbb{R}^{1\times d_\text{GNN}^{\ell-1}}$ is the aggregated information for node $x$ at the $\ell$-th GNN layer. Since $x$ belongs to a node in either $\mathcal{U}$ or $\mathcal{V}$, $\text{AGGREGATE}_{x}^{\ell}$ aggregates feature information of connected items if $x$ is a user node, and vice versa. In the update step, we use $\text{UPDATE}_{x}^{\ell}$ to obtain the $\ell$-th embedding vector ${\bf h}_{x}^{\ell}$ from the aggregated information ${\bf m}_{x}^{\ell}$ as follows:
\begin{align}
\label{update_}
{\bf h}_{x}^{\ell} \leftarrow \text{UPDATE}_{x}^{\ell}\big(x,{\bf m}_{x}^{\ell}\big).
\end{align}
We note that, for each node $x$, we randomly initialize the learnable 0-th embeddings (i.e., ${\bf h}^0_{x}$) due to the fact that we have no side information for users and items in our setting as in \cite{he2020lightgcn,chen2020revisiting}. Additionally, we present another function in our GNN model, namely $\text{LAYER-AGG}_x^{L_{\text{GNN}}}$, which performs {\em layer aggregation} similarly as in \cite{xu2018representation}. This operation is motivated by the argument that {\em oversmoothing} tends to occur in GNN-based recommender systems if the last GNN layer's embedding vectors are used as the final embedding ${\bf Z}^p$ \cite{li2018deeper}. To alleviate the oversmoothing problem, we calculate the embedding vector ${\bf z}_x^p\in \mathbb{R}^{1\times d}$ of node $x$ via layer aggregation as follows:

\begin{align}
\label{layer_agg}
    {\bf z}^p_x \leftarrow {\text{LAYER-AGG}}_x^{L_{\text{GNN}}} \left( {\big\{{\bf h}_x^{\ell}\big\}}_{\ell=0}^{\ell=L_{\text{GNN}}}\right),
\end{align}
which results in the embeddings ${\bf Z}^p$ for the graph $G^p$ where $L_\text{GNN}$ is the number of GNN layers. The GNN architecture of our \textsf{SiReN} method including the above three functions $\text{AGGREGATE}_x^{\ell}$, $\text{UPDATE}_x^{\ell}$, and $\text{LAYER-AGG}_x^{L_{\text{GNN}}}$ is illustrated in Fig. \ref{fig:GNN_}.

\begin{rmk}
    \label{GNN_reco_ex}
    Now, let us state how the above three functions in (\ref{aggregate_})--(\ref{layer_agg}) can be specified by several types of GNN-based recommender systems. As one state-of-the-art method, \color{black} NGCF \cite{wang2019neural} can be implemented by using 
    \begin{subequations}
    \begin{align}
        \text{AGGREGATE}_{1,x}^{\ell} = &\text{ }W_{\text{GNN};1}^{\ell} {\bf h}_{x}^{\ell -1}\\
        \text{AGGREGATE}_{2,x}^{\ell}=&\sum_{y \in \mathcal{N}_{x}}\frac{1}{\sqrt{|\mathcal{N}_x||\mathcal{N}_y|}}\Big( W_{\text{GNN};1}^{\ell}{\bf h}_y^{\ell -1}\\
        \nonumber
        &+W_{\text{GNN};2}^{\ell}({\bf h}_x^{\ell -1}\odot {\bf h}_y^{\ell -1}) \Big) \\
        \text{UPDATE}_x^{\ell} = &\text{ }\text{LeakyReLU}({\bf m}_{1,x}^{\ell}+{\bf m}_{2,x}^{\ell}) \\
        {\text{LAYER-AGG}}_x^{L_{\text{GNN}}} = &\text{ }{\bf h}_x^0 {\big| \big|} {\bf h}_x^1 {\big| \big|} \cdot \cdot \cdot {\big| \big|}  {\bf h}_x^{L_{\text{GNN}}},
    \end{align}
    \end{subequations}
    where $W_{\text{GNN};1}^{\ell},~ W_{\text{GNN};2}^{\ell}\in\mathbb{R}^{d_\text{GNN}^{\ell-1} \times d_\text{GNN}^{\ell}}$ are learnable weight transformation matrices, ${\big| \big|}$ is the concatenation operator, $L_{\text{GNN}}$ is the number of GNN layers, and $\text{LeakyReLU}$ is an activation function with a parameter $\alpha >0$:
    \begin{align}
        \text{LeakyReLU}(x)=\begin{cases}
                        x \text{ if $x>0$} \\
                        \alpha x \text{ otherwise}.
                    \end{cases}
    \end{align}
    \color{black}
    In addition, as another state-of-the-art method for recommendation, LR-GCCF \cite{chen2020revisiting} can be implemented by using
    \begin{subequations}
    \begin{align}
		\text{AGGREGATE}_x^{\ell}&=\sum_{y\in \mathcal{N}_x\cup \{x\}}\dfrac{1}{\sqrt{|\mathcal{N}_x|+1}\sqrt{|\mathcal{N}_y|+1}}{\bf h}_y^{\ell-1} \\
		\text{UPDATE}_x^{\ell}&={\bf m}^\ell_x\cdot W^{\ell}_{\text{GNN}} \\
		{\text{LAYER-AGG}}_x^{L_{\text{GNN}}} &= {\bf h}_x^0 {\big| \big|} {\bf h}_x^1 {\big| \big|} \cdot \cdot \cdot {\big| \big|}  {\bf h}_x^{L_{\text{GNN}}},
	\end{align}
	where $W_{\text{GNN}}^{\ell}\in\mathbb{R}^{d_\text{GNN}^{\ell-1} \times d_\text{GNN}^{\ell}}$ is a learnable weight transformation matrix. \color{black}As the most recently developed state-of-the-art method, \color{black}LightGCN \cite{he2020lightgcn} can be specified according to the following function setting:
	\end{subequations}
	\begin{subequations}
	\begin{align}
			\text{AGGREGATE}_x^{\ell}&=\sum_{y\in \mathcal{N}_x}\dfrac{1}{\sqrt{|\mathcal{N}_x|}\sqrt{|\mathcal{N}_y|}}{\bf h}^{\ell-1}_y \\
			\text{UPDATE}_x^{\ell}&={\bf m}^\ell_x\\
			\text{LAYER-AGG}_x^{L_{\text{GNN}}} &=\dfrac{1}{L_{\text{GNN}}+1}\sum_{\ell=0}^{L_{\text{GNN}}}{{\bf h}_x^{\ell}}.
	\end{align}
	\end{subequations}
\end{rmk}

Second, we pay our attention to the MLP architecture designed for the graph $G^n$ having negative edges. We calculate the embeddings ${\bf Z}^n$ using the MLP as follows:
\begin{subequations}
\begin{align}
{\bf Z}^n_{\ell} &= \text{ReLU}\big( {\bf Z}_{\ell-1}^nW^{\ell}_{\text{MLP}}+{\bf 1}_{\text{MLP}}{\bf b}^{\ell}_{\text{MLP}}  \big) \\
{\bf Z}^n &=  {\bf Z}^n_{L_{\text{MLP}}},
\end{align}
\end{subequations}
where $\ell=1,2,\cdots,L_{\text{MLP}}$; $L_{\text{MLP}}$ is the number of MLP layers; $\text{ReLU}(x)=\max{(0,x)}$; $W_{\text{MLP}}^{\ell}\in\mathbb{R}^{d_\text{MLP}^{\ell-1} \times d_\text{MLP}^{\ell}}$ is a learnable weight transformation matrix; ${\bf b}_{\text{MLP}}^{\ell}\in \mathbb{R}^{1 \times d_\text{MLP}^{\ell}}$ is a bias vector; $d_\text{MLP}^{\ell}$ is the dimension of the latent representation vector ${\bf Z}_{\ell}^n$ at the $\ell$-th MLP layer; ${\bf 1}_{\text{MLP}}\in \mathbb{R}^{(M+N)\times 1}$ is the all-ones vector; and ${\bf Z}_0^n\in \mathbb{R}^{(M+N)\times d_\text{MLP}^{0}}$ is the learnable 0-th layer's embedding matrix for all nodes in the set $\mathcal{U}\cup \mathcal{V}$, which is randomly initialized. That is, ${\big\{W_{\text{MLP}}^{\ell}\big\}}_{\ell=1}^{\ell=L_{\text{MLP}}}$, ${\big\{{\bf b}_{\text{MLP}}^{\ell}\big\}}_{\ell=1}^{\ell=L_{\text{MLP}}}$, and ${\bf Z}_0^n$ correspond to the model parameters of $\text{MLP}_{\theta_2}$ in (\ref{align_MLP_neg}).

Third, we turn to describing the attention model. The importance $({\bf \alpha}^p, {\bf \alpha}^n)$ in (\ref{align_attn}) represents the attention values of two embeddings ${\bf Z}^p$ and ${\bf Z}^n$ for all nodes in $\mathcal{U}\cup \mathcal{V}$. Let us focus on node $x\in \mathcal{U}\cup \mathcal{V}$ whose embedding vectors calculated for the graphs $G^p$ and $G^n$ are given by ${\bf z}^p_{x},{\bf z}^n_{x}\in \mathbb{R}^{1\times d}$, respectively. Let $w_x^p$ and $w_x^n$ denote attention values of the two embeddings ${\bf z}^p_x$ and ${\bf z}^n_x$, respectively, for node $x$. Then, our attention model learns a weight transformation matrix $W_{attn}\in \mathbb{R}^{d' \times d}$, an attention vector ${\bf q}\in \mathbb{R}^{d' \times 1}$, and a bias vector ${\bf b}\in \mathbb{R}^{d' \times 1}$ with a dimension $d'$, corresponding to the model parameters of $\text{ATTENTION}_{\theta_3}$ in (\ref{align_attn}), as follows:
\begin{subequations}
\begin{align}
\label{weight_attn1}
w_x^p&={\bf q}^T tanh(W_{attn} {{\bf z}^p_x}^T + {\bf b}) \\
\label{weight_attn2}
w_x^n&={\bf q}^T tanh(W_{attn} {{\bf z}^n_x}^T + {\bf b}),
\end{align}
\end{subequations}
where $tanh(x)=\frac{\exp(x)-\exp(-x)}{\exp(x)+\exp(-x)}$ is the hyperbolic tangent activation function. By normalizing the attention values in (\ref{weight_attn1}) and (\ref{weight_attn2}) according to the softmax function, we have
\begin{subequations}
\begin{align}
    \alpha_{x}^p&=\frac{\text{exp}({w_x^p})}{\text{exp}(w_x^p)+\text{exp}(w_x^n)}\\
    \alpha_{x}^n&=\frac{\text{exp}({w_x^n})}{\text{exp}(w_x^p)+\text{exp}(w_x^n)},
\end{align}
\end{subequations}
where $\alpha_{x}^p$ and $\alpha^n_{x}$ are the resulting importance of two embeddings ${\bf z}_x^p$ and ${\bf z}_x^n$, respectively, which thus yields the final embedding ${\bf z}_x = \alpha_{x}^p{\bf z}_x^p + \alpha_{x}^n{\bf z}_x^n$ for each node $x$ in (\ref{attn_final_emb}). 


\subsection{Optimization}
In this subsection, we explain the optimization of \textsf{SiReN} method in the training phase via our proposed loss function. We start by randomly initializing the learnable model parameters $\Theta=\{\theta_1, \theta_2, \theta_3\}$ in (\ref{align_GNN_pos})--(\ref{align_attn}) (refer to line 1 in Algorithm \ref{algorithm_main}). To train our learning models (i.e., the GNN, MLP, and attention models), we use \color{black}a training set \color{black} $\mathcal{D}_S$ consisting of multiple samples of a triplet $(u,i,j)$, where $(u,i,w_{ui}^s)\in\mathcal{E}^s$ and $j\in\mathcal{V}$ is a negative sample (i.e., an unobserved item), which is not in the set of direct neighbors of user $u$ in the signed bipartite graph $G^s$ (refer to line 6). \color{black}For sampling $j\in\mathcal{V}$, we use the degree-based noise distribution $P(j)\propto d_j^{3/4}$ \cite{mikolovskipgram,tang2015line}, where $d_j$ is the degree of item $j$. \color{black}More specifically, when $N_{\text{neg}}$ denotes the number of negative samples, we first acquire the set of edges, $\mathcal{E}^s$, in $G^s$ and then sample $N_{\text{neg}}$ negative samples $\{j_n\}_{n=1}^{N_{\text{neg}}}$ for each $(u,i,w_{ui}^s)\in\mathcal{E}^s$ \color{black} using the distribution $P(j_n)$ \color{black}in order to create new samples of a node triplet $(u,i,j_n)$ for all $n\in\{1,\cdots,N_{\text{neg}}\}$, which yields \color{black}the training set \color{black} $\mathcal{D}_S$ as follows:

	\begin{align}
	\label{data_batch}
	\mathcal{D}_S=\big\{(u,i,j_n)\big|(u,i,w_{ui}^s)\in\mathcal{E}^s,j_n\notin \mathcal{N}_u,n\in\{1,\cdots,N_{\text{neg}}\}  \big\},
	\end{align}
where $\mathcal{N}_u$ is the set of neighbor nodes of user $u$ in $G^s$. We further subsample \color{black}mini-batches \color{black}$\mathcal{D}_S'\subset\mathcal{D}_S$ to efficiently train our learning models (refer to line 7). The sampled triplets are fed into the loss function in the training loop along with the calculated embeddings $\bf Z$ for all nodes in $\mathcal{U}\cup \mathcal{V}$ (refer to lines 9--12). 

Now, we present our {\em sign-aware} BPR loss function, which is built upon the original BPR loss \cite{BPRMF09} widely used in recommender systems (see \cite{zheng2018spectral, wu2019neural, wu2020joint, wu2020diffnet++, wang2019neural, chen2020revisiting, he2020lightgcn} and references therein). To this end, we define a user $u$'s {\em predicted preference} for an item $i$ as the inner product of user and item final embeddings:
	\begin{align}
    \hat{r}_{ui} \triangleq {\bf z}_u{\bf z}_i^T,
	\end{align}
which is used for establishing our loss function.
	
However, simply employing the original BPR loss is not desirable in our setting since it is a pairwise loss based on the relative order between observed and unobserved items by basically assuming that high ratings are more reflective of a user's preferences with higher prediction values of $\hat{r}_{ui}$ than the case of unobserved ones. On the other hand, our objective function should account for two types of observed items, which include both positive and negative relations between users and items, as well as unobserved ones. \color{black}The proposed sign-aware BPR loss function is designed in such a way that the predicted preference for an observed item is higher than its unobserved counterparts along with the induced difference between high and low ratings (i.e., strong and weak user--item bonds).

To this end, more formally, we define a ternary relation $>_u(i,j,w)\triangleq \{(i,j,w)|\hat{r}_{ui}>\hat{r}_{uj}~\text{if}~w>0~\text{and}~\hat{r}_{ui}>{1 \over c}\hat{r}_{uj}~\text{otherwise}\} \subset\mathcal{V}\times \mathcal{V} \times (\mathbb{R}\setminus \{0\})$, where $c>1$ is a hyperparameter used for adjusting two levels of user preferences for observed items during training. In the relation, a higher value of $c$ indicates a larger gap between the degrees of user preferences for observed items. \color{black}Based on the relation $>_u$, we aim at minimizing the following loss function $\mathcal{L}$ for a given \color{black}mini\color{black}-batch $\mathcal{D}_S'$ with the $L_2$ regularization:



\color{black}
	\begin{align}
	\label{lossfunction}
		\mathcal{L}=\mathcal{L}_0+\lambda_{\text{reg}}\norm{\Theta}^2,
	\end{align}
where $\lambda_{\text{reg}}$ is a hyperparameter that controls the $L_2$ regularization strength; $\Theta$ represents the model parameters; and $\mathcal{L}_0$ is the sign-aware BPR loss term realized by
	\begin{align}
    \label{sign_BPRloss}
    \mathcal{L}_0=-\sum_{(u,i,j)\in\mathcal{D}_S'}{\log{p(>_u(i,j,w^s_{ui})\big| \Theta)}}.
    \end{align}
\color{black}Here, to capture the above relation $>_u(i,j,w)$ for each $(u,i,j)\in\mathcal{D}_S'$, we model the likelihood in (\ref{sign_BPRloss}) as

    \begin{align}
        \label{likelihood} 
        p(>_u(i,j,w^s_{ui})\big| \Theta)=\begin{cases}
                        \sigma \big(\hat{r}_{ui}-\hat{r}_{uj}\big) \text{ if $w^s_{ui}>0$} \\
                        \sigma \big(\text{c}\cdot\hat{r}_{ui}-\hat{r}_{uj}\big) \text{ otherwise,}
                    \end{cases}
    \end{align}
where $\sigma(x)=\frac{1}{1+\exp(-x)}$ is the sigmoid function and $c>1$. Our model is trained through the loss function in (\ref{lossfunction}) towards the objective of (i) $\hat{r}_{ui}>\hat{r}_{uj}$ for pairs $(u,i)$ with high ratings and (ii) $\hat{r}_{ui}>\frac{1}{c}\hat{r}_{uj}$ for pairs $(u,i)$ with low ratings. In consequence, it is possible to more elaborately learn representations of nodes depending on both positive and negative relations. 
\color{black}

\section{Experimental Evaluation}
In this section, we first describe real-world datasets used in the evaluation. We also present five competing methods including two baseline MF methods and three state-of-the-art GNN-based methods for comparison. After describing performance metrics and our experimental settings, we comprehensively evaluate the performance of our \textsf{SiReN} method and five benchmark methods. The source code for \textsf{SiReN} can be accessed via https://github.com/woni-seo/SiReN-reco.

\subsection{Datasets}
We conduct experiments on three real-world datasets, which are widely adopted for evaluating recommender systems. For all experiments, we use user--item interactions with ratings in each dataset as the input. The main statistics of each dataset, including the number of users, the number of items, the number of ratings, the density, and the rating scale, are summarized in Table \ref{data_stat}. In the following, we explain important characteristics of the datasets briefly.

\textbf{MovieLens-1M (ML-1M)\footnote{{https://grouplens.org/datasets/movielens/1m/.}}}. This is the most popular dataset in movie recommender systems, which consists of 5-star ratings (i.e., integer values from 1 to 5) of movies given by users \cite{harper2015movielens}.

\textbf{Amazon-Book\footnote{{https://jmcauley.ucsd.edu/data/amazon/index.html.}}}. Among the Amazon-Review dataset containing product reviews and metadata, we select the Amazon-Book dataset, which consists of 5-star ratings \cite{he2016ups}. We remove users/items that have less than 20 interactions similarly as in \cite{he2016vbpr}.

\textbf{Yelp\footnote{{https://www.yelp.com/dataset.}}}. This dataset is a local business review data consisting of 5-star ratings. As in the Amazon-Book dataset, we remove users/items that have less than 20 interactions.

\begin{table}[t]
    \centering
    \begin{tabular}{lccc} 
        \toprule
        \textbf{Dataset} & \textbf{ML-1M}& \textbf{Amazon-Book} & \textbf{Yelp}\\
        \midrule
        \# of users ($M$) & 6,040 & 35,736 & 41,772 \\
        \# of items ($N$) & 3,952 & 38,121 & 30,037 \\
        \# of ratings & 1,000,209 & 1,960,674&2,116,215\\
        Density (\%) &4.19&0.14&0.16 \\
        Rating scale &1--5&1--5&1--5\\
        \bottomrule
    \end{tabular}
    \caption{Statistics of three real-world datasets.}
    \label{data_stat}
\end{table}

\subsection{Benchmark Methods}
In this subsection, we present two baseline MF methods and \textcolor{black}{four} state-of-the-art GNN methods for comparison.

\textbf{BPRMF} \cite{BPRMF09}. This baseline method is a MF model optimized by the BPR loss, which assumes that each user prefers the items with which he/she has interacted to items with no interaction.

\textbf{NeuMF} \cite{he2017neural}. As another popular baseline, NeuMF is a neural CF model, which generalizes standard MF and uses multiple hidden layers to generate user and item embeddings.

\textbf{NGCF} \cite{wang2019neural}. This state-of-the-art GNN-based approach follows basic operations inherited from the standard GCN \cite{kipf2017semi} to explore the high-order connectivity information. More specifically, NGCF stacks embedding layers and concatenates embeddings obtained in all layers to constitute the final embeddings.

\textbf{LR-GCCF} \cite{chen2020revisiting}. LR-GCCF is a state-of-the-art GCN-based CF model. As two main characteristics, this model uses only linear transformation without nonlinear activation and concatenates all layers' embeddings to alleviate oversmoothing at deeper layers \cite{li2018deeper}.

\textbf{LightGCN} \cite{he2020lightgcn}. LightGCN simplifies the design of GCN \cite{kipf2017semi} to make the model more appropriate for recommendation by including only the most essential component such as neighborhood aggregation without nonlinear activation and weight transformation operations. Similarly as in LR-GCCF, this approach uses the weighted sum of embeddings learned at all layers as the final embedding.

\color{black}
\textbf{SGCN} \cite{derr2018signed}. SGCN is a generalized version of GCN \cite{kipf2017semi} and harnesses the balance theory while capturing both positive and negative edges coherently in the aggregation process. Although SGCN was primarily aimed at conducting the link sign prediction task in signed unipartite graphs, it can also be applied to signed bipartite graphs. 

\color{black}

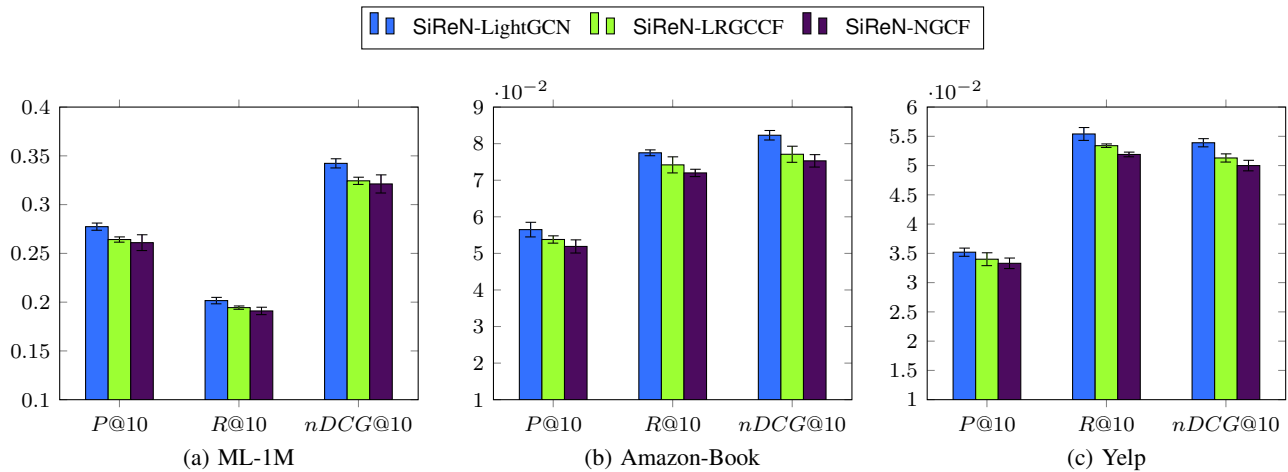
\begin{figure*}
\centering
\begin{tikzpicture}
\begin{groupplot}[
ybar=0pt,
enlarge x limits=0.25,
legend columns=-1,
legend pos=south east,
footnotesize,
symbolic x coords={$P@10$,$R@10$,$nDCG@10$},
xtick=data,
ymin=0.1, ymax=0.4, width=0.35\textwidth,
nodes near coords align={vertical},
group style={
group size=3 by 1,
xlabels at=edge bottom,
ylabels at=edge left,
xticklabels at=edge bottom}]

\nextgroupplot[bar width=0.3cm, xlabel={(a) ML-1M},legend style = { column sep = 5pt, legend columns = -1, legend to name = grouplegend},
    legend to name=grouplegend,
        legend entries={
            \textsf{SiReN}-LightGCN, \textsf{SiReN}-LRGCCF, \textsf{SiReN}-NGCF
        },
]
    \addplot[style={black,fill=bblue,mark=none}, error bars/.cd,
y dir=both,y explicit]
        coordinates {($P@10$, 0.2774)+-(0.0037,0.0037) ($R@10$,0.2016)+-(0.0033,0.0033) ($nDCG@10$,0.3423)+-(0.0047,0.0047)};
    \addlegendentry{\textsf{SiReN}-LightGCN}
    \addplot[style={black,fill=ggreen,mark=none}, error bars/.cd,
y dir=both,y explicit]
        coordinates {($P@10$, 0.2642)+-(0.0027,0.0027) ($R@10$,0.1944)+-(0.0017,0.0017) ($nDCG@10$,0.3244)+-(0.0037,0.0037)};
    \addlegendentry{\textsf{SiReN}-LRGCCF}
    \addplot[style={black,fill=ppurple,mark=none}, error bars/.cd,
y dir=both,y explicit]
         coordinates {($P@10$, 0.261)+-(0.0081,0.0081) ($R@10$,0.191)+-(0.0038,0.0038) ($nDCG@10$,0.3212)+-(0.0093,0.0093)};
    \addlegendentry{\textsf{SiReN}-NGCF}
\nextgroupplot[bar width=0.3cm, xlabel={(b) Amazon-Book},ymax=0.09,ymin=0.01]
    \addplot[style={black,fill=bblue,mark=none}, error bars/.cd,
y dir=both,y explicit]
        coordinates {($P@10$, 0.0565)+-(0.002,0.002) ($R@10$,0.0775)+-(0.0008,0.0008) ($nDCG@10$,0.0823)+-(0.0013,0.0013)};
    \addplot[style={black,fill=ggreen,mark=none}, error bars/.cd,
y dir=both,y explicit]
        coordinates {($P@10$, 0.0538)+-(0.001,0.001) ($R@10$,0.0742)+-(0.0022,0.0022) ($nDCG@10$,0.0771)+-(0.0022,0.0022)};
    \addplot[style={black,fill=ppurple,mark=none}, error bars/.cd,
y dir=both,y explicit]
         coordinates {($P@10$, 0.0519)+-(0.0018,0.0018) ($R@10$,0.072)+-(0.001,0.001) ($nDCG@10$,0.0753)+-(0.0017,0.0017)};
\nextgroupplot[bar width=0.3cm, xlabel={(c) Yelp},ymax=0.06,ymin=0.01]
    \addplot[style={black,fill=bblue,mark=none}, error bars/.cd,
y dir=both,y explicit]
        coordinates {($P@10$, 0.0352)+-(0.0007,0.0007) ($R@10$,0.0554)+-(0.0011,0.0011) ($nDCG@10$,0.0539)+-(0.0007,0.0007)};
    \addplot[style={black,fill=ggreen,mark=none}, error bars/.cd,
y dir=both,y explicit]
        coordinates {($P@10$, 0.034)+-(0.0011,0.0011) ($R@10$,0.0534)+-(0.0003,0.0003) ($nDCG@10$,0.0513)+-(0.0007,0.0007)};
    \addplot[style={black,fill=ppurple,mark=none}, error bars/.cd,
y dir=both,y explicit]
         coordinates {($P@10$, 0.0333)+-(0.0009,0.0009) ($R@10$,0.0519)+-(0.0004,0.0004) ($nDCG@10$,0.05)+-(0.0009,0.0009)};
\end{groupplot}
\node at ($(group c2r1) + (0, 3cm) $) {\ref{grouplegend}};
\end{tikzpicture}
\caption{\color{black}Performance comparison according to different GNN models \color{black} in our \textsf{SiReN} method for each dataset.}
\label{Fig_RQ1}
\end{figure*}
\subsection{Performance Metrics}
To validate the performance of the proposed \textsf{SiReN} method and the six benchmark methods, we adopt three metrics, which are widely used to evaluate the accuracy of top-$K$ recommendation. Let $Te_u^+$ and $R_u (K)$ denote the ground truth set (i.e., the set of items rated by user $u$ in the test set) and the top-$K$ recommendation list for user $u$, respectively. In the following, we describe each of metrics for recommendation accuracy.

The precision $P@K$ is defined as the ratio of relevant items to the set of recommended items and is expressed as
\begin{align}
P@K=\dfrac{1}{M}\sum_{u\in \mathcal{U}}{\dfrac{\big|Te_u^+\cap R_u(K) \big|}{K}}.
\end{align}
The recall $R@K$ is defined as the ratio of relevant items to the ground truth set and is expressed as
\begin{align}
R@K=\dfrac{1}{M}\sum_{u\in \mathcal{U}}{\dfrac{\big| Te_u^+ \cap R_u(K) \big|}{\big| Te_u^+ \big|}}.
\end{align}
The normalized discounted cumulative gain $nDCG@K$ \cite{NDCG} measures a ranking quality of the recommendation list by assigning higher scores to relevant items at top-$K$ ranking positions in the list:
\begin{align}
	nDCG@K = \dfrac{1}{M}\sum_{u\in \mathcal{U}}{nDCG_u@K}.
\end{align}
Let $y_k$ be the binary relevance of the $k$-th item $i_k$ in $R_u(K)$ for each user $u$: $y_k=1$ if $i_k \in Te_u^+$ and $0$ otherwise. Then, $nDCG_u@K$ can be computed as
\begin{align}
nDCG_u@K&=\dfrac{DCG_u@K}{IDCG_u@K},
\end{align}
where $DCG_u@K$ is
\begin{align}
DCG_u@K&=\sum_{i=1}^{K}{\dfrac{2^{y_i}-1}{\log_2(i+1)}}
\end{align}
and $IDCG_u@K$ indicates the ideal case of $DCG_u@K$ (i.e., all relevant items are at the top rank in $R_u(K)$). Note that all metrics are in a range of $[0,1]$, and higher values represent better performance.

\subsection{Experimental Setup}
In this subsection, we describe the experimental settings of neural networks in our \textsf{SiReN} method. We implement \textsf{SiReN} via PyTorch Geometric \cite{fey2019pyg}, which is a geometric deep learning extension library in PyTorch. In our experiments, we adopt GNN models for the graph $G^p$ with positive edges and the 2-layer MLP architecture for the graph $G^n$ with negative edges. We use the Xavier initializer \cite{xavier_} to initialize the model parameters $\Theta = \{\theta_1, \theta_2, \theta_3\}$. We use dropout regularization \cite{srivastava2014dropout} with the probability of 0.5 for the MLP and attention models in (\ref{align_MLP_neg}) and (\ref{align_attn}). We set the dimension of the embedding space and all hidden latent spaces to 64; the number of negative samples, $N_{\text{neg}}$, to 40; \color{black} the hyperparameter $c$ in (\ref{likelihood}) to 2;\footnote{\color{black}It was empirically found that the case of $c>2$ does not lead to high performance regardless of datasets even if it is not presented in this article for the sake of brevity.\color{black}} \color{black}and the strength of $L_2$ regularization, $\lambda_{\text{reg}}$, to 0.1 for the ML-1M dataset and 0.05 for the Amazon-Book and Yelp datasets. We train our model using the Adam optimizer \cite{kingma2015adam} with a learning rate of 0.005. 

For each dataset, we conduct 5-fold cross-validation by splitting it into two subsets: 80\% of the ratings (i.e., user--item interactions) as the training set and 20\% of the ratings as the test set. In the training set, when we implement five benchmark methods, we regard only the items with the rating scores of 4 and 5 as observed interactions by removing the ratings whose scores are lower than 4 as in \cite{chen2019collaborative, wu2019neural,wu2020diffnet++}; however, when we implement our \textsf{SiReN} method, we utilize all user--item interactions including low ratings in the training set while the parameter $w_o$, indicating the design criterion for signed bipartite graph construction, is set to 3.5.\footnote{Our empirical findings reveal that such a setting in \textsf{SiReN} consistently leads to superior performance to that of other values of $w_o$ for all datasets having the 1--5 rating scale. \textcolor{black}{This is obvious due to the fact that, in our experiments, the rating scores of 4 and 5 correspond to positive user--item interactions as well as the ground truth set.}} It is worthwhile to note that, for fair comparison, the test set consists of only the ratings of 4 and 5 as the {\em ground truth} set for all the methods including \textsf{SiReN}.

\subsection{Experimental Results}
In this subsection, our empirical study is designed to answer the following four key research questions.
\begin{itemize}
	\item [\textbf{RQ1}.] How do underlying GNN models affect the performance of the \textsf{SiReN} method?
	\item [\textbf{RQ2}.] Which model architecture is appropriate to the graph $G^n$ with negative edges?
	\item [\textbf{RQ3}.] How much does the \textsf{SiReN} method improve the top-$K$ recommendation over baseline and state-of-the-art methods? 
	\item [\textbf{RQ4}.] How robust is our \textsf{SiReN} method with respect to interaction sparsity levels?
\end{itemize}
To answer these research questions, we comprehensively carry out experiments in the following.

\paragraph{Comparative Study Among GNN Models Used for $G^p$ (RQ1)}
In Fig. \ref{Fig_RQ1}, for all datasets, we evaluate the accuracy of top-$K$ recommendation in terms of $P@K$, $R@K$, and $nDCG@K$ when $K$ is set to 10 while using various GNN models used for the graph $G^p$ in our \textsf{SiReN} method. Since our method is GNN-model-agnostic, any existing GNN models can be adopted; however, in our experiments, we adopt three state-of-the-art GNN models that exhibit superior performance in recommender systems from the literature, namely NGCF \cite{wang2019neural} (\textsf{SiReN}-NGCF), LR-GCCF \cite{chen2020revisiting} (\textsf{SiReN}-LRGCCF), and LightGCN \cite{he2020lightgcn} (\textsf{SiReN}-LightGCN). From Fig. \ref{Fig_RQ1}, we observe that \textsf{SiReN}-LightGCN consistently outperforms other models for all performance metrics. As discussed in \cite{he2020lightgcn}, this is because nonlinear activation and weight transformation operations in GNNs rather tend to degrade the recommendation accuracy; LightGCN thus attempted to simplify the design of GCN by removing such operations. It turns out that such a gain achieved by LightGCN is also possible in our \textsf{SiReN} model that contains three learning models including the GNN, MLP, and attention models.

From these findings, we use \textsf{SiReN}-LightGCN in our subsequent experiments unless otherwise stated.

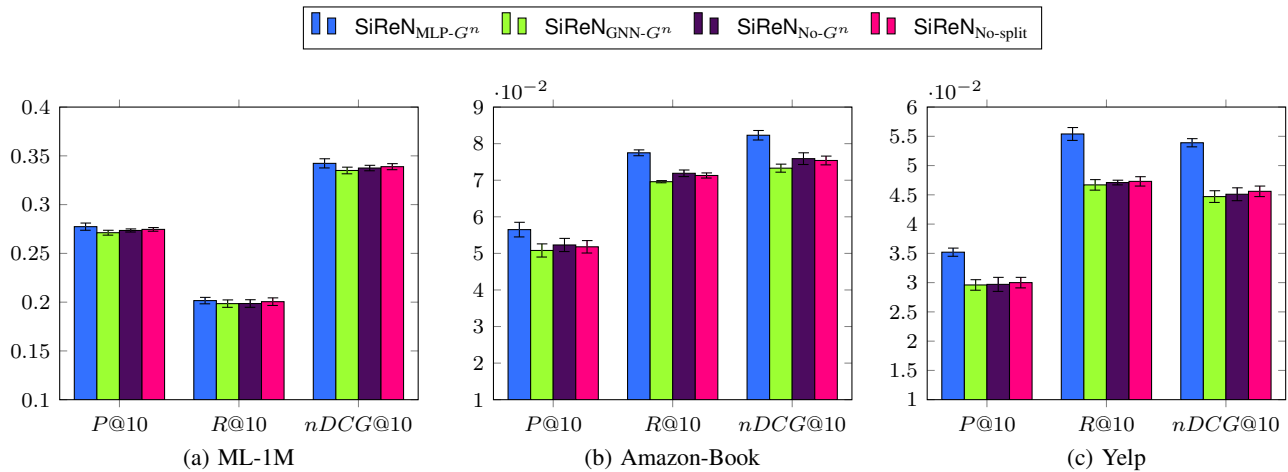
\begin{figure*}
\centering
\begin{tikzpicture}
\begin{groupplot}[
ybar=0pt,
enlarge x limits=0.25,
legend columns=-1,
legend pos=south east,
footnotesize,
symbolic x coords={$P@10$,$R@10$,$nDCG@10$},
xtick=data,
ymin=0.1, ymax=0.4, width=0.35\textwidth,
nodes near coords align={vertical},
group style={
group size=3 by 1,
xlabels at=edge bottom,
ylabels at=edge left,
xticklabels at=edge bottom}]

\nextgroupplot[bar width=0.3cm, xlabel={(a) ML-1M},legend style = { column sep = 5pt, legend columns = -1, legend to name = grouplegend2},
    legend to name=grouplegend2,
    legend entries={$\textsf{SiReN}_{\text{MLP-}G^n}$,$\textsf{SiReN}_{\text{GNN-}G^n}$,$\textsf{SiReN}_{\text{No-}G^n}$
    }, 
]
    \addplot[style={black,fill=bblue,mark=none}, error bars/.cd,
y dir=both,y explicit]
        coordinates {($P@10$, 0.2774)+-(0.0037,0.0037) ($R@10$,0.2016)+-(0.0033,0.0033) ($nDCG@10$,0.3423)+-(0.0047,0.0047)};
    \addlegendentry{$\textsf{SiReN}_{\text{MLP-}G^n}$}
    \addplot[style={black,fill=ggreen,mark=none}, error bars/.cd,
y dir=both,y explicit]
        coordinates {($P@10$, 0.2712)+-(0.0025,0.0025) ($R@10$,0.1985)+-(0.0038,0.0038) ($nDCG@10$,0.335)+-(0.0034,0.0034)};
    \addlegendentry{$\textsf{SiReN}_{\text{GNN-}G^n}$}
    \addplot[style={black,fill=ppurple,mark=none}, error bars/.cd,
y dir=both,y explicit]
         coordinates {($P@10$, 0.2735)+-(0.0016,0.0016) ($R@10$,0.1986)+-(0.0039,0.0039) ($nDCG@10$,0.3375)+-(0.0028,0.0028)};
    \addlegendentry{$\textsf{SiReN}_{\text{No-}G^n}$}
    \addplot[style={black,fill=ppink,mark=none}, error bars/.cd,
y dir=both,y explicit]
         coordinates {($P@10$, 0.2746)+-(0.0019,0.0019) ($R@10$,0.2005)+-(0.0039,0.0039) ($nDCG@10$,0.3389)+-(0.0031,0.0031)};
    \addlegendentry{$\textsf{SiReN}_{\text{\textcolor{black}{No-split}}}$}
\nextgroupplot[bar width=0.3cm, xlabel={(b) Amazon-Book},ymax=0.09,ymin=0.01]
    \addplot[style={black,fill=bblue,mark=none}, error bars/.cd,
y dir=both,y explicit]
        coordinates {($P@10$, 0.0565)+-(0.002,0.002) ($R@10$,0.0775)+-(0.0008,0.0008) ($nDCG@10$,0.0823)+-(0.0013,0.0013)};
    \addplot[style={black,fill=ggreen,mark=none}, error bars/.cd,
y dir=both,y explicit]
        coordinates {($P@10$, 0.0508)+-(0.0018,0.0018) ($R@10$,0.0696)+-(0.0003,0.0003) ($nDCG@10$,0.0733)+-(0.0011,0.0011)};
    \addplot[style={black,fill=ppurple,mark=none}, error bars/.cd,
y dir=both,y explicit]
         coordinates {($P@10$, 0.0523)+-(0.0018,0.0018) ($R@10$,0.0719)+-(0.0009,0.0009) ($nDCG@10$,0.0759)+-(0.0016,0.0016)};
    \addplot[style={black,fill=ppink,mark=none}, error bars/.cd,
y dir=both,y explicit]
         coordinates {($P@10$, 0.0518)+-(0.0017,0.0017) ($R@10$,0.0713)+-(0.0007,0.0007) ($nDCG@10$,0.0754)+-(0.0012,0.0012)};
\nextgroupplot[bar width=0.3cm, xlabel={(c) Yelp},ymax=0.06,ymin=0.01]
    \addplot[style={black,fill=bblue,mark=none}, error bars/.cd,
y dir=both,y explicit]
        coordinates {($P@10$, 0.0352)+-(0.0007,0.0007) ($R@10$,0.0554)+-(0.0011,0.0011) ($nDCG@10$,0.0539)+-(0.0007,0.0007)};
    \addplot[style={black,fill=ggreen,mark=none}, error bars/.cd,
y dir=both,y explicit]
        coordinates {($P@10$, 0.0296)+-(0.0009,0.0009) ($R@10$,0.0467)+-(0.0009,0.0009) ($nDCG@10$,0.0447)+-(0.001,0.001)};
    \addplot[style={black,fill=ppurple,mark=none}, error bars/.cd,
y dir=both,y explicit]
         coordinates {($P@10$, 0.0297)+-(0.0012,0.0012) ($R@10$,0.0471)+-(0.0004,0.0004) ($nDCG@10$,0.0451)+-(0.0011,0.0011)};
    \addplot[style={black,fill=ppink,mark=none}, error bars/.cd,
y dir=both,y explicit]
         coordinates {($P@10$, 0.03)+-(0.0009,0.0009) ($R@10$,0.0473)+-(0.0008,0.0008) ($nDCG@10$,0.0456)+-(0.0009,0.0009)};

\end{groupplot}
\node at ($(group c2r1) + (0, 3cm) $) {\ref{grouplegend2}};
\end{tikzpicture}
\caption{\color{black}Performance comparison according to several model architectures used for the graph $G^n$ in our \textsf{SiReN} method \textcolor{black}{for each dataset}.}
\label{Fig_RQ3}
\end{figure*}

\paragraph{Comparative Study Among Model Architectures Used for $G^n$ (RQ2)}
We perform another comparative study \textcolor{black}{four} among model architectures used for the graph $G^n$ with negative edges. We adopted the MLP for $G^n$ since negative edges in $G^n$ can undermine the assortativity and thus message passing to such dissimilar nodes would not be feasible. In this experiment, we empirically validate this claim by taking into account \textcolor{black}{three} other design scenarios. We evaluate the accuracy of top-$K$ recommendation when $K$ is set to 10 for all datasets. First, we recall the original \textsf{SiReN} method employing MLP for $G^n$, dubbed $\textsf{SiReN}_{\text{MLP-}G^n}$.

Second, instead of employing MLP, we introduce $\textsf{SiReN}_{\text{GNN-}G^n}$ that uses an additional GNN model for the graph $G^n$ to calculate the embedding vectors ${\bf Z}^n$:
\begin{align}
    {\bf Z}^n = \text{GNN}_{\theta_1'}(G^n),
\end{align}
where $\theta_1'$ is the learned model parameters of $\text{GNN}$ for the graph $G^n$. In this experiment, we adopt LightGCN \cite{he2020lightgcn} among GNN models.

Third, as an ablation study, we introduce $\textsf{SiReN}_{\text{No-}G^n}$ that does not calculate the embedding vectors ${\bf Z}^n$. In other words, we only use the embedding vectors ${\bf Z}^p$ in (\ref{align_GNN_pos}) as the final embeddings ${\bf Z}$ in (\ref{attn_final_emb}) (i.e., ${\bf Z} = {\bf Z}^p$). Note that $\textsf{SiReN}_{\text{No-}G^n}$ is identical to the model architecture of LightGCN \cite{he2020lightgcn}, which generates embedding vectors by aggregating only the information of positively connected neighbors. However, unlike LightGCN, $\textsf{SiReN}_{\text{No-}G^n}$ utilizes the sign-aware BPR loss in (\ref{lossfunction})--(\ref{likelihood}) as our objective function in the process of optimization. 

\color{black} \textcolor{black}{Fourth, we present $\textsf{SiReN}_\text{No-split}$, a variant of \textsf{SiReN}} that does not partition the signed bipartite graph $G^s$ into the two edge-disjoint graphs. That is, \textcolor{black}{$\textsf{SiReN}_\text{No-split}$ generates the embeddings ${\bf Z}$ in (\ref{attn_final_emb}) based on} the model architecture of LightGCN \cite{he2020lightgcn} without distinguishing \textcolor{black}{positive and negative edges}. However, \textcolor{black}{unlike LightGCN, $\textsf{SiReN}_\text{No-split}$ adopts the sign-aware BPR loss in the optimization.} \color{black}

From Fig. \ref{Fig_RQ3}, our findings are as follows:
\begin{itemize}
    \item $\textsf{SiReN}_{\text{MLP-}G^n}$ is always superior to $\textsf{SiReN}_{\text{GNN-}G^n}$ regardless of \textcolor{black}{datasets and }performance metrics, which indeed validates our claim addressed in Remark \ref{rmk_1}.
    \color{black}
    \item \textcolor{black}{$\textsf{SiReN}_{\text{MLP-}G^n}$ is always superior to $\textsf{SiReN}_\text{No-split}$ regardless of datasets and performance metrics. This implies that partitioning the signed bipartite graph into two edge-disjoint graphs can capture the user preferences more precisely.}
    \color{black}
    \item \textcolor{black}{$\textsf{SiReN}_{\text{GNN-}G^n}$ exhibits the worst performance among the four model architectures used for $G^n$ for all cases. This implies that message passing to dissimilar nodes in $G^n$ is not effective.} \color{black}
    \item Although all the models are \color{black} trained via our sign-aware BPR loss, both positive and negative relations in the signed bipartite graph $G^s$ are not precisely captured during training unless an appropriate model architecture is designed for the graph $G^n$.

\end{itemize}

From these findings, we use $\textsf{SiReN}_{\text{MLP-}G^n}$ in our subsequent experiments unless otherwise stated.

\begin{table*}[h!]
\centering
\setlength\tabcolsep{2.5pt} 
\scalebox{1}{
\begin{tabular}{ccccc|ccc|cccccc}
\toprule
\multicolumn{1}{c}{} & {} &  \multicolumn{3}{c|}{\color{black}$K=5$} & \multicolumn{3}{c|}{$K=10$}  & \multicolumn{3}{c}{$K=15$}  \\
 \cmidrule{3-11} 
  Dataset& Method & $P@K$ & $R@K$ & $nDCG@K$ & $P@K$ & $R@K$ & $nDCG@K$& $P@K$ & $R@K$ & $nDCG@K$ \\ 
\midrule
    \multirow{7}{*}{\rotatebox{90}{ML-1M}} 
    & BPRMF &  0.2360\tiny{$\pm$0.0058}  &  0.0746\tiny{$\pm$0.0017}  &  0.2536\tiny{$\pm$0.0081}  &  0.1999\tiny{$\pm$0.0044}  &  0.1227\tiny{$\pm$0.0019}  &  0.2363\tiny{$\pm$0.0063}  &  0.1772\tiny{$\pm$0.0033}  &  0.1608\tiny{$\pm$0.0016}  &  0.2314\tiny{$\pm$0.0053}  \\
    & NeuMF           &  0.2785\tiny{$\pm$0.0054}  &  0.0966\tiny{$\pm$0.0011}  &  0.299\tiny{$\pm$0.0035}  &  0.2397\tiny{$\pm$0.0030}  &  0.1599\tiny{$\pm$0.0011}  &  0.2847\tiny{$\pm$0.0035}  &  0.2138\tiny{$\pm$0.0027}  &  0.2098\tiny{$\pm$0.0023}  &  0.2827\tiny{$\pm$0.0032}  \\
    & NGCF            &  0.2973\tiny{$\pm$0.0043}  &  0.1099\tiny{$\pm$0.0026}  &  0.3238\tiny{$\pm$0.0045}  &  0.2477\tiny{$\pm$0.0023}  &  0.1748\tiny{$\pm$0.0025}  &  0.3031\tiny{$\pm$0.0033}  &  0.2174\tiny{$\pm$0.0022}  &  0.2229\tiny{$\pm$0.0027}  &  0.2985\tiny{$\pm$0.0029}  \\
    & LR-GCCF        &  0.3052\tiny{$\pm$0.0033}  &  0.114\tiny{$\pm$0.0023}  &  0.333\tiny{$\pm$0.0044}  &  0.2539\tiny{$\pm$0.0027}  &  0.1802\tiny{$\pm$0.0031}  &  0.3117\tiny{$\pm$0.0039}  &  0.2220\tiny{$\pm$0.0025}  &  0.2292\tiny{$\pm$0.0046}  &  0.3066\tiny{$\pm$0.0042}  \\
    & LightGCN       &  0.3218\tiny{$\pm$0.0022}  &  0.1206\tiny{$\pm$0.0011}  &  0.3519\tiny{$\pm$0.0023}  &  0.2679\tiny{$\pm$0.0013}  &  0.1909\tiny{$\pm$0.0016}  &  0.3297\tiny{$\pm$0.0018}  &  0.2349\tiny{$\pm$0.0016}  &  0.2432\tiny{$\pm$0.0029}  &  0.3249\tiny{$\pm$0.0022}  \\
    & \color{black}\textsf{SGCN}  & 
    \color{black}0.2484\tiny{$\pm$0.0033}  &  \color{black}0.091\tiny{$\pm$0.0017}  &  \color{black}0.2683\tiny{$\pm$0.0035}  & 
    \color{black}0.1873\tiny{$\pm$0.002}  &  \color{black}0.1492\tiny{$\pm$0.0026}  &  \color{black}0.2547\tiny{$\pm$0.0024}  & \color{black}0.1873\tiny{$\pm$0.002}  &  \color{black}0.1932\tiny{$\pm$0.0037}  &  \color{black}0.253\tiny{$\pm$0.0026}  & \\
    & \textsf{SiReN}  & 
    \color{black}\underline{0.3328}\tiny{$\pm$0.0054}  &  \color{black}\underline{0.1279}\tiny{$\pm$0.0027}  &  \color{black}\underline{0.3637}\tiny{$\pm$0.0055}  & 
    \color{black}\underline{0.2774}\tiny{$\pm$0.0037}  &  \color{black}\underline{0.2016}\tiny{$\pm$0.0033}  &  \color{black}\underline{0.3423}\tiny{$\pm$0.0047}  &  \color{black}\underline{0.2444}\tiny{$\pm$0.0025}  &  \color{black}\underline{0.2569}\tiny{$\pm$0.0029}  &  \color{black}\underline{0.3388}\tiny{$\pm$0.004}  & \\
\midrule
\multirow{7}{*}{\rotatebox{90}{Amazon-Book}}
    & BPRMF          &  0.0298\tiny{$\pm$0.0039}  &  0.0209\tiny{$\pm$0.0035}  &  0.0331\tiny{$\pm$0.0046}  &  0.0263\tiny{$\pm$0.0033}  &  0.0365\tiny{$\pm$0.0057}  &  0.0371\tiny{$\pm$0.0053}  &  0.0243\tiny{$\pm$0.0029}  &  0.0500\tiny{$\pm$0.0074}  &  0.0419\tiny{$\pm$0.0058}  \\
    & NeuMF          &  0.0402\tiny{$\pm$0.0021}  &  0.0271\tiny{$\pm$0.0007}  &  0.0448\tiny{$\pm$0.002}  &  0.0339\tiny{$\pm$0.0018}  &  0.0452\tiny{$\pm$0.0010}  &  0.0483\tiny{$\pm$0.0018}  &  0.0303\tiny{$\pm$0.0014}  &  0.0597\tiny{$\pm$0.0012}  &  0.0530\tiny{$\pm$0.0017}  \\
    & NGCF         &  0.0463\tiny{$\pm$0.0014}  &  0.032\tiny{$\pm$0.0007}  &  0.0518\tiny{$\pm$0.0011}    &  0.0391\tiny{$\pm$0.0014}  &  0.0532\tiny{$\pm$0.0008}  &  0.0562\tiny{$\pm$0.0010}  &  0.0403\tiny{$\pm$0.0127}  &  0.0706\tiny{$\pm$0.0007}  &  0.0618\tiny{$\pm$0.0009}  \\
    & LR-GCCF        &  0.0469\tiny{$\pm$0.0016}  &  0.0324\tiny{$\pm$0.0002}  &  0.0527\tiny{$\pm$0.0012} &  0.0399\tiny{$\pm$0.0014}  &  0.0544\tiny{$\pm$0.0005}  &  0.0574\tiny{$\pm$0.0009}  &  0.0357\tiny{$\pm$0.0013}  &  0.0721\tiny{$\pm$0.0004}  &  0.0631\tiny{$\pm$0.0008}  \\
    & LightGCN       &  0.0529\tiny{$\pm$0.0015}  &  0.0362\tiny{$\pm$0.0007}  &  0.0596\tiny{$\pm$0.0011} &  0.0443\tiny{$\pm$0.0013}  &  0.0595\tiny{$\pm$0.0008}  &  0.0638\tiny{$\pm$0.0007}  &  0.0393\tiny{$\pm$0.0011}  &  0.0781\tiny{$\pm$0.0008}  &  0.0698\tiny{$\pm$0.0007}  \\
    & \color{black}SGCN & 
    \color{black}0.039\tiny{$\pm$0.0023}  &  \color{black}0.0267\tiny{$\pm$0.0012}  &  \color{black}0.0433\tiny{$\pm$0.0024}  & 
    \color{black}0.0336\tiny{$\pm$0.0018}  &  \color{black}0.0454\tiny{$\pm$0.0018}  &  \color{black}0.0475\tiny{$\pm$0.0023}  & \color{black}0.0304\tiny{$\pm$0.0016}  &  \color{black}0.0609\tiny{$\pm$0.0021}  &  \color{black}0.0527\tiny{$\pm$0.0024}  & \\
    & \textsf{SiReN}  &  \color{black}\underline{0.0678}\tiny{$\pm$0.0025}  &  \color{black}\underline{0.0474}\tiny{$\pm$0.0005}  &  \color{black}\underline{0.0766}\tiny{$\pm$0.0002}  &  \color{black}\underline{0.0565}\tiny{$\pm$0.0025}  &  \color{black}\underline{0.0775}\tiny{$\pm$0.0008}  &  \color{black}\underline{0.0823}\tiny{$\pm$0.0013}  &  \color{black}\underline{0.0497}\tiny{$\pm$0.0018}  &  \color{black}\underline{0.1009}\tiny{$\pm$0.0011}  &  \color{black}\underline{0.0897}\tiny{$\pm$0.0011}\\
\midrule
\multirow{7}{*}{\rotatebox{90}{Yelp}} 
    & BPRMF          &  0.0124\tiny{$\pm$0.0013}  &  0.0096\tiny{$\pm$0.0009}  &  0.0137\tiny{$\pm$0.0013}&  0.0116\tiny{$\pm$0.0014}  &  0.0180\tiny{$\pm$0.0020}  &  0.0165\tiny{$\pm$0.0017}  &  0.0111\tiny{$\pm$0.0013}  &  0.0257\tiny{$\pm$0.0026}  &  0.0194\tiny{$\pm$0.0021}  \\
    & NeuMF          &  0.0199\tiny{$\pm$0.0011}  &  0.015\tiny{$\pm$0.0012}  &  0.0226\tiny{$\pm$0.0015} &  0.0174\tiny{$\pm$0.0009}  &  0.0262\tiny{$\pm$0.0017}  &  0.0254\tiny{$\pm$0.0015}  &  0.0159\tiny{$\pm$0.0007}  &  0.0358\tiny{$\pm$0.0019}  &  0.0288\tiny{$\pm$0.0016}  \\
    & NGCF           &  0.0285\tiny{$\pm$0.0012}  &  0.0226\tiny{$\pm$0.0007}  &  0.0329\tiny{$\pm$0.001} &  0.0243\tiny{$\pm$0.0009}  &  0.0383\tiny{$\pm$0.0010}  &  0.0368\tiny{$\pm$0.0010}  &  0.0219\tiny{$\pm$0.0007}  &  0.0515\tiny{$\pm$0.0008}  &  0.0413\tiny{$\pm$0.0009}  \\
    & LR-GCCF      &  0.0303\tiny{$\pm$0.0014}  &  0.024\tiny{$\pm$0.0006}  &  0.0351\tiny{$\pm$0.0013}  &  0.0258\tiny{$\pm$0.0010}  &  0.0405\tiny{$\pm$0.0009}  &  0.0392\tiny{$\pm$0.0011}  &  0.0232\tiny{$\pm$0.0008}  &  0.0543\tiny{$\pm$0.0010}  &  0.0439\tiny{$\pm$0.0012}  \\
    & LightGCN     &  0.0333\tiny{$\pm$0.0011}  &  0.0259\tiny{$\pm$0.0004}  &  0.0386\tiny{$\pm$0.0009}  &  0.0281\tiny{$\pm$0.0009}  &  0.0435\tiny{$\pm$0.0007}  &  0.0427\tiny{$\pm$0.0008}  &  0.0251\tiny{$\pm$0.0008}  &  0.0582\tiny{$\pm$0.0010}  &  0.0476\tiny{$\pm$0.0008}  \\
    & \color{black}SGCN & 
    \color{black}0.0293\tiny{$\pm$0.001}  &  \color{black}0.0226\tiny{$\pm$0.0006}  &  \color{black}0.0332\tiny{$\pm$0.001}  & 
   \color{black} 0.0256\tiny{$\pm$0.0008}  &  \color{black}0.0395\tiny{$\pm$0.0011}  &  \color{black}0.0377\tiny{$\pm$0.001}  & \color{black}0.0232\tiny{$\pm$0.0007}  &  \color{black}0.0538\tiny{$\pm$0.0019}  &  \color{black}0.0426\tiny{$\pm$0.0012}  & \\
    & \textsf{SiReN}  &  \color{black}\underline{0.042}\tiny{$\pm$0.009}  &  \color{black}\underline{0.0332}\tiny{$\pm$0.0005}  &  \color{black}\underline{0.0488}\tiny{$\pm$0.0007}  &  \color{black}\underline{0.0352}\tiny{$\pm$0.0007}  &  \color{black}\underline{0.0554}\tiny{$\pm$0.0011}  &  \color{black}\underline{0.0539}\tiny{$\pm$0.0007}  &  \color{black}\underline{0.0314}\tiny{$\pm$0.0006}  &  \color{black}\underline{0.0737}\tiny{$\pm$0.0012}  &  \color{black}\underline{0.06}\tiny{$\pm$0.0006}\\
\midrule
\end{tabular}}
\caption{\color{black}Performance comparison among \textsf{SiReN} and six benchmark methods in terms of three performance metrics (average $\pm$ standard deviation) when $K\in\{5,10,15\}$. Here, the best method for each case is highlighted using underlines. }
\label{Q4table}
\end{table*}

\begin{table*}[h!]
\centering
\setlength\tabcolsep{2.5pt} 
\scalebox{1}{
\begin{tabular}{cccccc|ccc|cccccc}
\toprule 
\multicolumn{1}{c}{} & {}&{} & \multicolumn{3}{c|}{\color{black}$K=5$} & \multicolumn{3}{c|}{$K=10$}  & \multicolumn{3}{c}{$K=15$}  \\
 \cmidrule{4-12} 
  Dataset&Group& Method & $P@K$ & $R@K$ & $nDCG@K$ & $P@K$ & $R@K$ & $nDCG@K$& $P@K$ & $R@K$ & $nDCG@K$ \\ 
\midrule
    \multirow{21}{*}{\rotatebox{90}{ML-1M}}&
    \multirow{7}{*}{\rotatebox{90}{\thead{$[0,20)$}}} 
    & BPRMF        &  0.0625\tiny{$\pm$0.0065}  &  0.0882\tiny{$\pm$0.0149}  &  0.0858\tiny{$\pm$0.0131}  &  0.0485\tiny{$\pm$0.0056}  &  0.1348\tiny{$\pm$0.0200}  &  0.1046\tiny{$\pm$0.0148}  &  0.0433\tiny{$\pm$0.0031}  &  0.1810\tiny{$\pm$0.0145}  &  0.1223\tiny{$\pm$0.0128}  \\
    && NeuMF          &  0.0842\tiny{$\pm$0.0063}  &  0.1185\tiny{$\pm$0.0122}  &  0.1158\tiny{$\pm$0.0122}  &  0.0675\tiny{$\pm$0.0050}  &  0.1873\tiny{$\pm$0.0182}  &  0.1449\tiny{$\pm$0.0147}  &  0.0576\tiny{$\pm$0.0036}  &  0.2397\tiny{$\pm$0.0204}  &  0.1653\tiny{$\pm$0.0149}  \\
    && NGCF          &  0.0987\tiny{$\pm$0.0077}  &  0.1384\tiny{$\pm$0.0157}  &  0.1376\tiny{$\pm$0.0142}  &  0.0760\tiny{$\pm$0.0040}  &  0.2141\tiny{$\pm$0.0126}  &  0.1679\tiny{$\pm$0.0129}  &  0.0638\tiny{$\pm$0.0015}  &  0.2660\tiny{$\pm$0.0100}  &  0.1884\tiny{$\pm$0.0101}  \\
    && LR-GCCF        &  0.1035\tiny{$\pm$0.0062}  &  0.1457\tiny{$\pm$0.0131}  &  0.1457\tiny{$\pm$0.0153}  &  0.0776\tiny{$\pm$0.0037}  &  0.2148\tiny{$\pm$0.0145}  &  0.1736\tiny{$\pm$0.0162}  &  0.0643\tiny{$\pm$0.0025}  &  0.2670\tiny{$\pm$0.0139}  &  0.1938\tiny{$\pm$0.0147}  \\
    && LightGCN     &  0.1070\tiny{$\pm$0.0047}  &  0.1508\tiny{$\pm$0.0107}  &  0.1515\tiny{$\pm$0.0104}  &  0.0832\tiny{$\pm$0.0029}  &  0.2326\tiny{$\pm$0.0116}  &  0.1847\tiny{$\pm$0.0108}  &  0.069\tiny{$\pm$0.0028}  &  0.2878\tiny{$\pm$0.0146}  &  0.2063\tiny{$\pm$0.0108}  \\
    && \color{black}SGCN &  \color{black}0.0823\tiny{$\pm$0.0079}  &  \color{black}0.1151\tiny{$\pm$0.0084}  &  \color{black}0.1126\tiny{$\pm$0.0096}  \color{black}&  0.0649\tiny{$\pm$0.0065}  &  \color{black}0.1789\tiny{$\pm$0.0158}  &  \color{black}0.1393\tiny{$\pm$0.0122}  &  \color{black}0.0541\tiny{$\pm$0.0051}  &  \color{black}0.224\tiny{$\pm$0.0177}  &  \color{black}0.1566\tiny{$\pm$0.0123}  \\
    && \textsf{SiReN}   & 
    \color{black}\underline{0.123}\tiny{$\pm$0.008}  &  \color{black}\underline{0.1755}\tiny{$\pm$0.0152}  &  \color{black}\underline{0.1756}\tiny{$\pm$0.0168}  & 
    \color{black}\underline{0.0933}\tiny{$\pm$0.0053}  &  \color{black}\underline{0.2606}\tiny{$\pm$0.0129}  &  \color{black}\underline{0.2106}\tiny{$\pm$0.0149}  & \color{black}\underline{0.077}\tiny{$\pm$0.002}  &  \color{black}\underline{0.3223}\tiny{$\pm$0.0051}  &  \color{black}\underline{0.2345}\tiny{$\pm$0.0121}  & \\
\cmidrule{2-12} 
&\multirow{7}{*}{\rotatebox{90}{\thead{$[20,50)$}}}
    & BPRMF         &  0.0877\tiny{$\pm$0.0041}  &  0.0825\tiny{$\pm$0.0029}  &  0.1052\tiny{$\pm$0.0057}  &  0.0731\tiny{$\pm$0.0018}  &  0.1358\tiny{$\pm$0.0026}  &  0.1194\tiny{$\pm$0.0041}  &  0.0635\tiny{$\pm$0.0013}  &  0.1773\tiny{$\pm$0.0046}  &  0.1364\tiny{$\pm$0.0037}\\
    && NeuMF           &  0.1245\tiny{$\pm$0.0032}  &  0.1196\tiny{$\pm$0.0047}  &  0.1488\tiny{$\pm$0.004}&  0.1010\tiny{$\pm$0.0020}  &  0.1916\tiny{$\pm$0.0052}  &  0.1679\tiny{$\pm$0.0038}  &  0.0877\tiny{$\pm$0.0014}  &  0.2485\tiny{$\pm$0.0061}  &  0.1913\tiny{$\pm$0.0041}  \\
    && NGCF            &  0.1460\tiny{$\pm$0.0027}  &  0.1428\tiny{$\pm$0.0052}  &  0.1768\tiny{$\pm$0.0059}&  0.1149\tiny{$\pm$0.0021}  &  0.2198\tiny{$\pm$0.0073}  &  0.1961\tiny{$\pm$0.0060}  &  0.0959\tiny{$\pm$0.0015}  &  0.2727\tiny{$\pm$0.0086}  &  0.2177\tiny{$\pm$0.0061}  \\
    && LR-GCCF        &  0.1548\tiny{$\pm$0.0036}  &  0.1504\tiny{$\pm$0.0041}  &  0.1885\tiny{$\pm$0.0047} &  0.1193\tiny{$\pm$0.0026}  &  0.2298\tiny{$\pm$0.0050}  &  0.2062\tiny{$\pm$0.0048}  &  0.0993\tiny{$\pm$0.0022}  &  0.2844\tiny{$\pm$0.0077}  &  0.2284\tiny{$\pm$0.0055} \\
    && LightGCN      &  0.1633\tiny{$\pm$0.0027}  &  0.1596\tiny{$\pm$0.0037}  &  0.2003\tiny{$\pm$0.0037} &  0.1262\tiny{$\pm$0.0020}  &  0.2424\tiny{$\pm$0.0044}  &  0.2194\tiny{$\pm$0.0037}  &  0.1055\tiny{$\pm$0.0015}  &  0.3013\tiny{$\pm$0.0055}  &  0.2434\tiny{$\pm$0.0036}  \\
    && \color{black}SGCN  &\color{black}  0.1204\tiny{$\pm$0.0029}  & \color{black} 0.1179\tiny{$\pm$0.0028}  &  \color{black}0.1451\tiny{$\pm$0.0031}  & \color{black} 0.0985\tiny{$\pm$0.0022}  &  \color{black}0.1915\tiny{$\pm$0.0056}  &  \color{black}0.1657\tiny{$\pm$0.0032}  &  \color{black}0.084\tiny{$\pm$0.0026}  &  \color{black}0.2421\tiny{$\pm$0.0084}  &  \color{black}0.1865\tiny{$\pm$0.0043}  \\
    && \textsf{SiReN}  & 
   \color{black} \underline{0.1728}\tiny{$\pm$0.0036}  & \color{black} \underline{0.1722}\tiny{$\pm$0.0044}  &  \color{black}\underline{0.2147}\tiny{$\pm$0.0046}  & 
   \color{black} \underline{0.1342}\tiny{$\pm$0.0026}  & \color{black} \underline{0.2613}\tiny{$\pm$0.0053}  &  \color{black}\underline{0.2361}\tiny{$\pm$0.0044}  & \color{black}\underline{0.1116}\tiny{$\pm$0.0018}  &  \color{black}\underline{0.322}\tiny{$\pm$0.0052}  & \color{black} \underline{0.261}\tiny{$\pm$0.0044}  & \\
\cmidrule{2-12} 
&\multirow{7}{*}{\rotatebox{90}{\thead{$[50,\infty)$}}} 
    & BPRMF          &  0.3201\tiny{$\pm$0.0079}  &  0.0698\tiny{$\pm$0.0019}  &  0.3372\tiny{$\pm$0.0099}&  0.2722\tiny{$\pm$0.0068}  &  0.1156\tiny{$\pm$0.0035}  &  0.3021\tiny{$\pm$0.0083}  &  0.2418\tiny{$\pm$0.0052}  &  0.1514\tiny{$\pm$0.0037}  &  0.2852\tiny{$\pm$0.0071}  \\
    && NeuMF          &  0.3673\tiny{$\pm$0.0091}  &  0.0841\tiny{$\pm$0.0028}  &  0.385\tiny{$\pm$0.0089} &  0.3194\tiny{$\pm$0.0055}  &  0.1430\tiny{$\pm$0.0033}  &  0.3514\tiny{$\pm$0.0066}  &  0.2862\tiny{$\pm$0.0043}  &  0.1893\tiny{$\pm$0.0037}  &  0.3355\tiny{$\pm$0.0059}  \\
    && NGCF          &  0.3853\tiny{$\pm$0.0061}  &  0.0922\tiny{$\pm$0.0025}  &  0.4085\tiny{$\pm$0.006} &  0.3247\tiny{$\pm$0.0040}  &  0.1505\tiny{$\pm$0.0031}  &  0.3649\tiny{$\pm$0.0052}  &  0.2875\tiny{$\pm$0.0039}  &  0.1962\tiny{$\pm$0.0037}  &  0.3459\tiny{$\pm$0.0052}  \\
    && LR-GCCF      &  0.3931\tiny{$\pm$0.0038}  &  0.0944\tiny{$\pm$0.0019}  &  0.4167\tiny{$\pm$0.0052} &  0.3321\tiny{$\pm$0.0033}  &  0.1544\tiny{$\pm$0.0030}  &  0.3730\tiny{$\pm$0.0046}  &  0.2930\tiny{$\pm$0.0031}  &  0.2006\tiny{$\pm$0.0032}  &  0.3531\tiny{$\pm$0.0043}  \\
    && LightGCN    &  0.4146\tiny{$\pm$0.0044}  &  0.1\tiny{$\pm$0.0024}  &  0.4402\tiny{$\pm$0.0052}&  0.3502\tiny{$\pm$0.0026}  &  0.1635\tiny{$\pm$0.0029}  &  0.3898\tiny{$\pm$0.0109}  &  0.3098\tiny{$\pm$0.0024}  &  0.2125\tiny{$\pm$0.0034}  &  0.3736\tiny{$\pm$0.0043}  \\
    && \color{black}SGCN &  \color{black}0.3227\tiny{$\pm$0.0055}  & \color{black} 0.0766\tiny{$\pm$0.0017}  &  \color{black}0.3393\tiny{$\pm$0.0064}  & \color{black} 0.2763\tiny{$\pm$0.0029}  & \color{black} 0.1273\tiny{$\pm$0.0023}  &  \color{black}0.3393\tiny{$\pm$0.0064}  & \color{black} 0.2472\tiny{$\pm$0.0033}  & \color{black} 0.1682\tiny{$\pm$0.003}  &  \color{black}0.2926\tiny{$\pm$0.0048}  \\
    && \textsf{SiReN}  &  
   \color{black} \underline{0.4258}\tiny{$\pm$0.0075}  &  \color{black}\underline{0.1033}\tiny{$\pm$0.0029}  &  \color{black}\underline{0.4497}\tiny{$\pm$0.0079}  & \color{black} \underline{0.3604}\tiny{$\pm$0.005}  &  \color{black}\underline{0.1688}\tiny{$\pm$0.0035}  & \color{black} \underline{0.3896}\tiny{$\pm$0.0278}  &  \color{black}\underline{0.321}\tiny{$\pm$0.0033}  & \color{black} \underline{0.2211}\tiny{$\pm$0.0035}  &  \color{black}\underline{0.3844}\tiny{$\pm$0.0055}\\
    \midrule
\multirow{21}{*}{\rotatebox{90}{Amazon-Book}}&\multirow{6}{*}{\rotatebox{90}{\thead{$[0,20)$}}} 
    & BPRMF        &  0.0181\tiny{$\pm$0.0027}  &  0.0263\tiny{$\pm$0.0052}  &  0.0239\tiny{$\pm$0.0042}  &  0.0157\tiny{$\pm$0.0021}  &  0.0455\tiny{$\pm$0.008}  &  0.033\tiny{$\pm$0.0054}  &  0.0142\tiny{$\pm$0.0019}  &  0.0619\tiny{$\pm$0.0103}  &  0.0393\tiny{$\pm$0.0062}\\
    && NeuMF          &  0.0247\tiny{$\pm$0.0022}  &  0.0333\tiny{$\pm$0.0014}  &  0.032\tiny{$\pm$0.0018}  &  0.0199\tiny{$\pm$0.0016}  &  0.0542\tiny{$\pm$0.0013}  &  0.042\tiny{$\pm$0.002}  &  0.0171\tiny{$\pm$0.0012}  &  0.0699\tiny{$\pm$0.0017}  &  0.0482\tiny{$\pm$0.0021}\\
    && NGCF          &  0.0288\tiny{$\pm$0.0016}  &  0.0398\tiny{$\pm$0.0014}  &  0.0377\tiny{$\pm$0.0012}  &  0.0235\tiny{$\pm$0.0012}  &  0.0649\tiny{$\pm$0.0019}  &  0.0497\tiny{$\pm$0.0014}  &  0.0204\tiny{$\pm$0.0011}  &  0.0848\tiny{$\pm$0.0015}  &  0.0576\tiny{$\pm$0.0013}\\
    && LR-GCCF        &  0.0292\tiny{$\pm$0.0017}  &  0.0402\tiny{$\pm$0.0007}  &  0.0382\tiny{$\pm$0.001}  &  0.0241\tiny{$\pm$0.0013}  &  0.0665\tiny{$\pm$0.0013}  &  0.0508\tiny{$\pm$0.0009}  &  0.0208\tiny{$\pm$0.0012}  &  0.0863\tiny{$\pm$0.001}  &  0.0586\tiny{$\pm$0.0009}\\
    && LightGCN      &  0.0318\tiny{$\pm$0.0015}  &  0.0438\tiny{$\pm$0.0012}  &  0.042\tiny{$\pm$0.0008}  &  0.0258\tiny{$\pm$0.0011}  &  0.0712\tiny{$\pm$0.0021}  &  0.0551\tiny{$\pm$0.0008}  &  0.0224\tiny{$\pm$0.0011}  &  0.0923\tiny{$\pm$0.0022}  &  0.0635\tiny{$\pm$0.0009}\\
    && \color{black}SGCN &\color{black}  0.0267\tiny{$\pm$0.0069}  &  \color{black}0.0317\tiny{$\pm$0.0038}  &  \color{black}0.0335\tiny{$\pm$0.0054}  & \color{black} 0.0225\tiny{$\pm$0.0063}  & \color{black} 0.053\tiny{$\pm$0.0054}  &  \color{black}0.0427\tiny{$\pm$0.003}  & \color{black} 0.0199\tiny{$\pm$0.0058}  &  \color{black}0.0701\tiny{$\pm$0.0063}  &  \color{black}0.0492\tiny{$\pm$0.0025}  \\
    && \textsf{SiReN}  &  
   \color{black} \underline{0.0435}\tiny{$\pm$0.0028}  & \color{black} \underline{0.06}\tiny{$\pm$0.0013}  &  \color{black}\underline{0.0575}\tiny{$\pm$0.001}  &  \color{black}\underline{0.0344}\tiny{$\pm$0.0024}  &  \color{black}\underline{0.0946}\tiny{$\pm$0.0013}  &  \color{black}\underline{0.0742}\tiny{$\pm$0.0014}  &  \color{black}\underline{0.0292}\tiny{$\pm$0.0019}  & \color{black} \underline{0.1208}\tiny{$\pm$0.0018}  &  \color{black}\underline{0.0845}\tiny{$\pm$0.0013}\\
\cmidrule{2-12} 
&\multirow{7}{*}{\rotatebox{90}{\thead{$[20,50)$}}}
    & BPRMF         &  0.0242\tiny{$\pm$0.0031}  &  0.0214\tiny{$\pm$0.0035}  &  0.0268\tiny{$\pm$0.0038}  &  0.0211\tiny{$\pm$0.0026}  &  0.0371\tiny{$\pm$0.0058}  &  0.032\tiny{$\pm$0.0046}  &  0.0194\tiny{$\pm$0.0023}  &  0.0508\tiny{$\pm$0.0077}  &  0.0383\tiny{$\pm$0.0055}\\
    && NeuMF           &  0.0337\tiny{$\pm$0.0021}  &  0.0281\tiny{$\pm$0.0004}  &  0.0374\tiny{$\pm$0.002}  &  0.0281\tiny{$\pm$0.0019}  &  0.0468\tiny{$\pm$0.0008}  &  0.0426\tiny{$\pm$0.0015}  &  0.0249\tiny{$\pm$0.0016}  &  0.0622\tiny{$\pm$0.0010}  &  0.0497\tiny{$\pm$0.0015}\\
    && NGCF            &  0.0386\tiny{$\pm$0.0019}  &  0.033\tiny{$\pm$0.0006}  &  0.043\tiny{$\pm$0.0014}  &  0.0322\tiny{$\pm$0.0016}  &  0.0547\tiny{$\pm$0.0007}  &  0.0494\tiny{$\pm$0.0008}  &  0.0286\tiny{$\pm$0.0014}  &  0.0727\tiny{$\pm$0.0009}  &  0.0576\tiny{$\pm$0.0009}\\
    && LR-GCCF        &  0.0394\tiny{$\pm$0.0016}  &  0.0336\tiny{$\pm$0.0005}  &  0.0442\tiny{$\pm$0.0011}  &  0.033\tiny{$\pm$0.0016}  &  0.056\tiny{$\pm$0.0007}  &  0.0507\tiny{$\pm$0.0007}  &  0.0265\tiny{$\pm$0.0013}  &  0.0744\tiny{$\pm$0.0006}  &  0.0591\tiny{$\pm$0.0007}\\
    && LightGCN      &  0.0441\tiny{$\pm$0.0017}  &  0.0375\tiny{$\pm$0.001}  &  0.0496\tiny{$\pm$0.0012}  &  0.0364\tiny{$\pm$0.0016}  &  0.0615\tiny{$\pm$0.0009}  &  0.0562\tiny{$\pm$0.0005}  &  0.0319\tiny{$\pm$0.0014}  &  0.0807\tiny{$\pm$0.0009}  &  0.065\tiny{$\pm$0.0005}\\
    && \color{black}SGCN & \color{black} 0.0324\tiny{$\pm$0.0024}  & \color{black} 0.0275\tiny{$\pm$0.0015}  &  \color{black}0.0358\tiny{$\pm$0.0025}  & \color{black} 0.0277\tiny{$\pm$0.0019}  & \color{black} 0.0469\tiny{$\pm$0.002}  &  \color{black}0.0417\tiny{$\pm$0.0022}  &  \color{black}0.0249\tiny{$\pm$0.0016}  &  \color{black}0.0632\tiny{$\pm$0.0025}  &  \color{black}0.0491\tiny{$\pm$0.0024}  \\
    && \textsf{SiReN}  &  
  \color{black}  \underline{0.0576}\tiny{$\pm$0.0024}  &  \color{black}\underline{0.0489}\tiny{$\pm$0.0011}  &  \color{black}\underline{0.0647}\tiny{$\pm$0.0018}  & \color{black} \underline{0.0476}\tiny{$\pm$0.0022}  &  \color{black}\underline{0.0805}\tiny{$\pm$0.0016}  & \color{black} \underline{0.0735}\tiny{$\pm$0.0006}  &  \color{black}\underline{0.0414}\tiny{$\pm$0.002}  &  \color{black}\underline{0.1049}\tiny{$\pm$0.0019}  &  \color{black}\underline{0.0846}\tiny{$\pm$0.0006}\\
\cmidrule{2-12} 
&\multirow{7}{*}{\rotatebox{90}{\thead{$[50,\infty)$}}} 
    & BPRMF          &  0.0551\tiny{$\pm$0.0067}  &  0.0143\tiny{$\pm$0.0019}  &  0.0574\tiny{$\pm$0.007}  &  0.0497\tiny{$\pm$0.0058}  &  0.0255\tiny{$\pm$0.0034}  &  0.0533\tiny{$\pm$0.0063}  &  0.0466\tiny{$\pm$0.0052}  &  0.0357\tiny{$\pm$0.0046}  &  0.0527\tiny{$\pm$0.0062}\\
    && NeuMF          &  0.0717\tiny{$\pm$0.0034}  &  0.0183\tiny{$\pm$0.0007}  &  0.0756\tiny{$\pm$0.0036}  &  0.0625\tiny{$\pm$0.0026}  &  0.0318\tiny{$\pm$0.0007}  &  0.0683\tiny{$\pm$0.003}  &  0.057\tiny{$\pm$0.0023}  &  0.0433\tiny{$\pm$0.0008}  &  0.0658\tiny{$\pm$0.0025}\\
    && NGCF          &  0.0827\tiny{$\pm$0.0016}  &  0.0215\tiny{$\pm$0.0003}  &  0.0875\tiny{$\pm$0.0019}  &  0.0718\tiny{$\pm$0.002}  &  0.0371\tiny{$\pm$0.0003}  &  0.0789\tiny{$\pm$0.0019}  &  0.0652\tiny{$\pm$0.0014}  &  0.0505\tiny{$\pm$0.0004}  &  0.0761\tiny{$\pm$0.0012}\\
    && LR-GCCF       &  0.0834\tiny{$\pm$0.0031}  &  0.0216\tiny{$\pm$0.0004}  &  0.0882\tiny{$\pm$0.0033}  &  0.0729\tiny{$\pm$0.0022}  &  0.0378\tiny{$\pm$0.0002}  &  0.0799\tiny{$\pm$0.0026}  &  0.0665\tiny{$\pm$0.0019}  &  0.0515\tiny{$\pm$0.0004}  &  0.0774\tiny{$\pm$0.002}\\
    && LightGCN    &  0.0958\tiny{$\pm$0.0026}  &  0.0249\tiny{$\pm$0.0003}  &  0.1019\tiny{$\pm$0.0032}  &  0.0823\tiny{$\pm$0.0021}  &  0.0423\tiny{$\pm$0.0003}  &  0.0911\tiny{$\pm$0.0025}  &  0.0747\tiny{$\pm$0.0018}  &  0.0574\tiny{$\pm$0.0004}  &  0.0877\tiny{$\pm$0.0018}\\
    && \color{black}SGCN  & \color{black} 0.0708\tiny{$\pm$0.0036}  &  \color{black}0.0185\tiny{$\pm$0.0007}  &  \color{black}0.0742\tiny{$\pm$0.0038}  &\color{black}  0.0624\tiny{$\pm$0.0035}  &  \color{black}0.0324\tiny{$\pm$0.0015}  &  \color{black}0.0677\tiny{$\pm$0.0037}  & \color{black} 0.0571\tiny{$\pm$0.003}  &  \color{black}0.0442\tiny{$\pm$0.002}  &  \color{black}0.0657\tiny{$\pm$0.0033}  \\
    && \textsf{SiReN}  &  
  \color{black}  \underline{0.1176}\tiny{$\pm$0.0046}  &  \color{black}\underline{0.0309}\tiny{$\pm$0.0006}  &  \color{black}\underline{0.1249}\tiny{$\pm$0.0055}  & \color{black} \underline{0.1008}\tiny{$\pm$0.0032}  &  \color{black}\underline{0.0525}\tiny{$\pm$0.0008}  &\color{black}  \underline{0.1117}\tiny{$\pm$0.004}  &  \color{black}\underline{0.0909}\tiny{$\pm$0.0028}  &  \color{black}\underline{0.0706}\tiny{$\pm$0.0011}  &  \color{black}\underline{0.1071}\tiny{$\pm$0.0033}\\
    \midrule
    \multirow{21}{*}{\rotatebox{90}{Yelp}}&\multirow{6}{*}{\rotatebox{90}{\thead{$[0,20)$}}} 
    & BPRMF       &  0.0066\tiny{$\pm$0.0009}  &  0.0107\tiny{$\pm$0.0012}  &  0.0091\tiny{$\pm$0.0013} &  0.0061\tiny{$\pm$0.0009}  &  0.0197\tiny{$\pm$0.0024}    &  0.0132\tiny{$\pm$0.0018}  &  0.0057\tiny{$\pm$0.0008}  &  0.0276\tiny{$\pm$0.0031}  &  0.0161\tiny{$\pm$0.0022}  \\
    && NeuMF         &  0.0097\tiny{$\pm$0.0011}  &  0.0155\tiny{$\pm$0.0018}  &  0.0138\tiny{$\pm$0.0016} &  0.0083\tiny{$\pm$0.0009}  &  0.0268\tiny{$\pm$0.0024}  &  0.0189\tiny{$\pm$0.0020}  &  0.0075\tiny{$\pm$0.0007}  &  0.0364\tiny{$\pm$0.0026}  &  0.0224\tiny{$\pm$0.0021}  \\
    && NGCF          &  0.0152\tiny{$\pm$0.0013}  &  0.0244\tiny{$\pm$0.0012}  &  0.0219\tiny{$\pm$0.0014} &  0.0128\tiny{$\pm$0.0010}  &  0.0413\tiny{$\pm$0.0017}  &  0.0295\tiny{$\pm$0.0016}  &  0.0114\tiny{$\pm$0.0009}  &  0.0552\tiny{$\pm$0.0014}  &  0.0347\tiny{$\pm$0.0016} \\
    && LR-GCCF        &  0.0162\tiny{$\pm$0.0018}  &  0.0259\tiny{$\pm$0.0014}  &  0.0232\tiny{$\pm$0.0017}&  0.0134\tiny{$\pm$0.0013}  &  0.0434\tiny{$\pm$0.0023}  &  0.0310\tiny{$\pm$0.0022}  &  0.0120\tiny{$\pm$0.0001}  &  0.0584\tiny{$\pm$0.0022}  &  0.0366\tiny{$\pm$0.0021}  \\
    && LightGCN      &  0.0171\tiny{$\pm$0.0012}  &  0.0274\tiny{$\pm$0.0009}  &  0.0248\tiny{$\pm$0.0012}&  0.0144\tiny{$\pm$0.0010}  &  0.0462\tiny{$\pm$0.0012}  &  0.0332\tiny{$\pm$0.0015}  &  0.0127\tiny{$\pm$0.0009}  &  0.0615\tiny{$\pm$0.0014}  &  0.0389\tiny{$\pm$0.0016}  \\
    && \color{black}SGCN  & \color{black} 0.0152\tiny{$\pm$0.0012}  &  \color{black}0.0243\tiny{$\pm$0.0014}  &  \color{black}0.0213\tiny{$\pm$0.0013}  &  \color{black}0.0131\tiny{$\pm$0.001}  & \color{black} 0.0422\tiny{$\pm$0.0027}  &  \color{black}0.0293\tiny{$\pm$0.0018}  & \color{black} 0.0119\tiny{$\pm$0.0009}  & \color{black} 0.0574\tiny{$\pm$0.0031}  &  \color{black}0.035\tiny{$\pm$0.002}  \\
    && \textsf{SiReN}   &  
 \color{black}   \underline{0.0224}\tiny{$\pm$0.0016}  & \color{black} \underline{0.0359}\tiny{$\pm$0.001}  &  \color{black}\underline{0.0321}\tiny{$\pm$0.0012}  & \color{black} \underline{0.0184}\tiny{$\pm$0.0012}  &  \color{black}\underline{0.0596}\tiny{$\pm$0.0016}  & \color{black} \underline{0.0427}\tiny{$\pm$0.0014}  &  \color{black}\underline{0.0163}\tiny{$\pm$0.0011}  &  \color{black}\underline{0.079}\tiny{$\pm$0.002}  &  \color{black}\underline{0.05}\tiny{$\pm$0.0016}\\
\cmidrule{2-12} 
&\multirow{7}{*}{\rotatebox{90}{\thead{$[20,50)$}}}
    & BPRMF        &  0.01\tiny{$\pm$0.0013}  &  0.0097\tiny{$\pm$0.0008}  &  0.0111\tiny{$\pm$0.0013} &  0.0094\tiny{$\pm$0.0014}  &  0.0183\tiny{$\pm$0.0019}  &  0.0146\tiny{$\pm$0.0016}  &  0.0090\tiny{$\pm$0.0012}  &  0.0263\tiny{$\pm$0.0025}  &  0.0181\tiny{$\pm$0.0019}  \\
    && NeuMF           &  0.0157\tiny{$\pm$0.001}  &  0.0152\tiny{$\pm$0.0011}  &  0.018\tiny{$\pm$0.0011}&  0.0137\tiny{$\pm$0.0008}  &  0.0267\tiny{$\pm$0.0017}  &  0.0222\tiny{$\pm$0.0013}  &  0.0124\tiny{$\pm$0.0008}  &  0.0365\tiny{$\pm$0.0020}  &  0.0265\tiny{$\pm$0.0015}  \\
    && NGCF            &  0.0235\tiny{$\pm$0.0012}  &  0.0232\tiny{$\pm$0.0008}  &  0.0271\tiny{$\pm$0.0009} &  0.0198\tiny{$\pm$0.0010}  &  0.0392\tiny{$\pm$0.0011}  &  0.0329\tiny{$\pm$0.0009}  &  0.0177\tiny{$\pm$0.0009}  &  0.0526\tiny{$\pm$0.0010}  &  0.0388\tiny{$\pm$0.0009} \\
    && LR-GCCF        &  0.0245\tiny{$\pm$0.0015}  &  0.0244\tiny{$\pm$0.0005}  &  0.0286\tiny{$\pm$0.0012} &  0.0208\tiny{$\pm$0.0011}  &  0.0413\tiny{$\pm$0.0007}  &  0.0348\tiny{$\pm$0.0008}  &  0.0186\tiny{$\pm$0.0009}  &  0.0552\tiny{$\pm$0.0010}  &  0.0409\tiny{$\pm$0.0009}  \\
    && LightGCN     &  0.0269\tiny{$\pm$0.0014}  &  0.0266\tiny{$\pm$0.0005}  &  0.0314\tiny{$\pm$0.0011}  &  0.0224\tiny{$\pm$0.0012}  &  0.0443\tiny{$\pm$0.0008}  &  0.0377\tiny{$\pm$0.0009}  &  0.0200\tiny{$\pm$0.0010}  &  0.0592\tiny{$\pm$0.0013}  &  0.0442\tiny{$\pm$0.0009}  \\
    && \color{black}SGCN  &\color{black}  0.0237\tiny{$\pm$0.0015}  & \color{black} 0.0231\tiny{$\pm$0.0006}  &  \color{black}0.027\tiny{$\pm$0.0012}  & \color{black} 0.0207\tiny{$\pm$0.0012}  & \color{black} 0.0405\tiny{$\pm$0.0009}  &  \color{black}0.0335\tiny{$\pm$0.001}  & \color{black} 0.0187\tiny{$\pm$0.0009}  &\color{black}  0.0549\tiny{$\pm$0.0019}  &  \color{black}0.0398\tiny{$\pm$0.0011}  \\
    && \textsf{SiReN}  &  
   \color{black} \underline{0.0344}\tiny{$\pm$0.0012}  & \color{black} \underline{0.0339}\tiny{$\pm$0.0007}  &  \color{black}\underline{0.0402}\tiny{$\pm$0.0007}  & \color{black} \underline{0.0287}\tiny{$\pm$0.0009}  &  \color{black}\underline{0.0567}\tiny{$\pm$0.0015}  & \color{black} \underline{0.0482}\tiny{$\pm$0.0006}  &  \color{black}\underline{0.0254}\tiny{$\pm$0.0008}  &  \color{black}\underline{0.0753}\tiny{$\pm$0.0016}  &  \color{black}\underline{0.0564}\tiny{$\pm$0.0004}\\
\cmidrule{2-12} 
&\multirow{7}{*}{\rotatebox{90}{\thead{$[50,\infty)$}}} 
    & BPRMF          &  0.0256\tiny{$\pm$0.0021}  &  0.008\tiny{$\pm$0.0005}  &  0.0263\tiny{$\pm$0.0017}&  0.0241\tiny{$\pm$0.0024}  &  0.0150\tiny{$\pm$0.0015}  &  0.0256\tiny{$\pm$0.0020}  &  0.0236\tiny{$\pm$0.0025}  &  0.0221\tiny{$\pm$0.0024}  &  0.0269\tiny{$\pm$0.0024}  \\
    && NeuMF         &  0.0435\tiny{$\pm$0.0024}  &  0.0137\tiny{$\pm$0.0008}  &  0.0456\tiny{$\pm$0.0029} &  0.0384\tiny{$\pm$0.0018}  &  0.0240\tiny{$\pm$0.0014}  &  0.0420\tiny{$\pm$0.0023}  &  0.0352\tiny{$\pm$0.0013}  &  0.0331\tiny{$\pm$0.0014}  &  0.0423\tiny{$\pm$0.0020}  \\
    && NGCF         &  0.0584\tiny{$\pm$0.0023}  &  0.0187\tiny{$\pm$0.0004}  &  0.0617\tiny{$\pm$0.0023} &  0.0506\tiny{$\pm$0.0016}  &  0.0323\tiny{$\pm$0.0004}  &  0.0562\tiny{$\pm$0.0017}  &  0.0458\tiny{$\pm$0.0012}  &  0.0439\tiny{$\pm$0.0003}  &  0.0562\tiny{$\pm$0.0012}  \\
    && LR-GCCF    &  0.0629\tiny{$\pm$0.0018}  &  0.0201\tiny{$\pm$0.0004}  &  0.0671\tiny{$\pm$0.002}   &  0.0542\tiny{$\pm$0.0012}  &  0.0346\tiny{$\pm$0.0003}  &  0.0608\tiny{$\pm$0.0014} &  0.0492\tiny{$\pm$0.0013}  &  0.047\tiny{$\pm$0.0006}  &  0.0606\tiny{$\pm$0.0013}   \\
    && LightGCN    &  0.0699\tiny{$\pm$0.0015}  &  0.0222\tiny{$\pm$0.0004}  &  0.0748\tiny{$\pm$0.0015}&  0.0599\tiny{$\pm$0.0013}  &  0.0380\tiny{$\pm$0.0004}  &  0.0675\tiny{$\pm$0.0012}  &  0.0540\tiny{$\pm$0.0013}  &  0.0514\tiny{$\pm$0.0004}  &  0.0670\tiny{$\pm$0.0008}  \\
    && \color{black}SGCN   & \color{black} 0.0616\tiny{$\pm$0.0008}  &\color{black}  0.0194\tiny{$\pm$0.0004}  &  \color{black}0.0645\tiny{$\pm$0.0011}& \color{black} 0.054\tiny{$\pm$0.0008}  & \color{black} 0.0341\tiny{$\pm$0.0008}  &  \color{black}0.0593\tiny{$\pm$0.0009}  & \color{black} 0.0491\tiny{$\pm$0.0009}  &  \color{black}0.0465\tiny{$\pm$0.0011}  &  \color{black}0.0593\tiny{$\pm$0.0009}  \\
    && \textsf{SiReN}  &  
  \color{black}  \underline{0.0864}\tiny{$\pm$0.001}  & \color{black} \underline{0.0277}\tiny{$\pm$0.0004}  &  \color{black}\underline{0.0924}\tiny{$\pm$0.0009}  & \color{black} \underline{0.0732}\tiny{$\pm$0.0011}  &  \color{black}\underline{0.047}\tiny{$\pm$0.001}  & \color{black} \underline{0.0829}\tiny{$\pm$0.0009}  &  \color{black}\underline{0.0656}\tiny{$\pm$0.0008}  &\color{black}  \underline{0.0629}\tiny{$\pm$0.0013}  &  \color{black}\underline{0.082}\tiny{$\pm$0.0008}\\
\midrule
\end{tabular}}
\caption{\color{black}Performance comparison among \textsf{SiReN} and six benchmark methods in terms of three performance metrics (average $\pm$ standard deviation) according to three different user groups \color{black} when $K\in \{5, 10, 15\}$\color{black}. Here, the best method for each case is highlighted using underlines.}
\label{cold_table}
\end{table*}

\paragraph{Comparison With Benchmark Methods (RQ3)}
The performance comparison between our \textsf{SiReN} method and \textcolor{black}{four} state-of-the-art GNN methods, including NGCF \cite{wang2019neural}, LR-GCCF \cite{chen2020revisiting}\color{black}, LightGCN \cite{he2020lightgcn}, and SGCN \cite{derr2018signed} \color{black} as well as two baseline MF methods, including BPRMF \cite{BPRMF09} and NeuMF \cite{he2017neural}, for top-$K$ recommendation is comprehensively presented in Table \ref{Q4table} with respect to three performance metrics using three real-world datasets, where \color{black} $K\in\{5,10,15\}$. \color{black} We note that the hyperparameters in all the aforementioned benchmark methods are tuned differently according to each dataset so as to provide the best performance. In Table \ref{Q4table}, the value with an underline indicates the best performer for each case. We would like to make the following insightful observations:

\begin{itemize}
	\item Our \textsf{SiReN} method consistently outperforms five benchmark methods for all datasets regardless of the performance metrics and the values of $K$. The superiority of our method comes from the fact that we are capable of more precisely representing users' preferences without any information loss. This implies that low ratings are indeed informative as long as the low rating information is well exploited through judicious model design and optimization.
	\color{black}
	\item The second best performer is LightGCN for all the cases. The performance gap between our \textsf{SiReN} method ($X$) and the second best performer ($Y$) is the largest when the Amazon-Book dataset is used; the maximum improvement rates of 28.16\%, 30.93\%, and 28.52\% are achieved in terms of $P@5$, $R@5$, and $nDCG@5$, respectively, where the improvement rate ($\%$) is given by $\frac{X-Y}{Y}\times 100$. We recall that the Amazon-Book dataset has the lowest density (i.e., the highest sparsity) out of three datasets (refer to Table \ref{data_stat}). Thus, from the above empirical finding, it is seen that exploiting negative user--item interactions in sparser datasets would be more beneficial and effective in improving the recommendation accuracy.
	\item SGCN is \textcolor{black}{far inferior to some of state-of-the-art GNN methods (e.g., LR-GCCF and LightGCN). This implies that performance of the GNN model built upon the balance theory is not effective compared to other competing GNN-based recommender systems.}
	\item Two baseline MF methods reveal worse performance than that of three state-of-the-art GNN methods including NGCF, LR-GCCF, and LightGCN. This indicates that exploring the high-order connectivity information via GNNs indeed significantly improves the recommendation accuracy.
	\color{black}
	\item The gain of LightGCN over LR-GCCF and NGCF is consistently observed. We note that, in contrast to LR-GCCF and NGCF, LightGCN aggregates the information of neighbors without both nonlinear activation and weight transformation operations. Our experimental results coincide with the argument in \cite{he2020lightgcn} that a simple aggregator of the information of neighbors using a weighted sum is the most effective as long as GNN-based recommender systems are associated.

\end{itemize}

\paragraph{Robustness to Interaction Sparsity Levels (RQ4)}
Needless to say, the sparsity issue is one of crucial challenges on designing recommender systems since few user--item interactions are insufficient to generate high-quality embeddings \cite{nguyen2018npe,wang2019neural}. In this experiment, we demonstrate that making use of low rating scores for better representing users’ preferences enables us to alleviate this sparsity issue. To this end, we partition the set of users in the test set into three groups according to the number of interactions in the training set as in \cite{nguyen2018npe,wang2019neural}. More precisely, for each dataset, we split the users into three groups, each of which is composed of the users whose number of interactions in the training set ranges between $[0,20)$, $[20,50)$, and $[50,\infty)$, respectively. In Table \color{black} \ref{cold_table}, we comprehensively carry out the performance comparison between our \textsf{SiReN} method and six benchmark methods with respect to three performance metrics of top-$K$ recommendation using the ML-1M, Amazon-Book, and Yelp datasets, where experimental results are shown according to three interaction sparsity levels for each dataset. Our findings can be summarized as follows:  


\begin{itemize}
\color{black}
\item Our \textsf{SiReN} method consistently outperforms \textcolor{black}{all} the benchmark \textcolor{black}{method. This demonstrates} that exploiting the set of negative interactions in designing GNN-based recommender systems \textcolor{black}{is useful in} improving recommendation accuracy \textcolor{black}{regardless of interaction sparsity levels}.
\item The performance gap between our \textsf{SiReN} method and the second best performer is the largest for the user group having $[0,20)$ interactions in the Amazon-Book dataset; the maximum improvement rates of 36.79\%, 36.98\%, and 36.9\% are achieved in terms of $P@5$, $R@5$, and $nDCG@5$, respectively. As stated above, the performance improvement of \textsf{SiReN} over competing methods is significant when sparse datasets are used.
\color{black}
\item As the number of interactions per user increases, the performance is likely to be enhanced for all the methods regardless of types of datasets. It is obvious that more user--item interactions yield higher recommendation accuracy.
\end{itemize}
\color{black}
\section{Concluding Remarks}
In this paper, we explored a fundamentally important problem of how to take advantage of both high and low rating scores in developing GNN-based recommender systems. To tackle this challenge, we introduced a novel method, termed \textsf{SiReN}, that is designed based on sign-aware learning and optimization models along with a GNN architecture. Specifically, we presented an approach to 1) constructing a signed bipartite graph $G^s$ to distinguish users' positive and negative feedback and then partitioning $G^s$ into two edge-disjoint graphs $G^p$ and $G^n$ with positive and negative edges each, respectively, 2) generating two embeddings for $G^p$ and $G^n$ via a GNN model and an MLP, respectively, and then using an attention model to discover the final embeddings, and 3) training our learning models by establishing a sign-aware BPR loss function that captures each relation of positively and negatively connected neighbors. Using three real-world datasets, we demonstrated that our \textsf{SiReN} method remarkably outperforms \color{black} four \color{black} state-of-the-art GNN methods as well as two baseline MF methods while showing gains over the second best performer (i.e., LightGCN) by up to \color{black} 30.93\% \color{black} in terms of the recommendation accuracy. \color{black} We also demonstrated that our proposed method is robust to more challenging situations according to interaction sparsity levels by investigating that the performance improvement of \textsf{SiReN} over state-of-the-art methods is significant when sparse datasets are used. Additionally, we empirically showed the effectiveness of MLP used for the graph $G^n$ with negative edges.

Potential avenues of future research include the design of a more sophisticated GNN model that fits well into $G^n$ in signed bipartite graphs. Here, the challenges lie in developing a new information aggregation and propagation mechanism. \color{black}Furthermore, the design of multi-criteria recommender systems utilizing sign awareness remains for future work. \color{black}


%



\section*{Acknowledgments}
The work of W.-Y. Shin was supported by the National Research Foundation of Korea (NRF) grant funded by the Korea government (MSIT) (No. 2021R1A2C3004345) and by the Yonsei University Research Fund of 2021 (2021-22-0083). The work of S. Lim was supported by the Institute of Information \& Communications Technology Planning \& Evaluation (IITP) grant funded by the Korea Government (MSIT) (No. 2020-0-01441, Artifical Intelligence Convergence Research Center (Chungnam National University)). The authors would like to thank Dr. Hajoon Ko from Harvard University for his helpful comments.

\ifCLASSOPTIONcaptionsoff
  \newpage
\fi



\bibliographystyle{IEEEtran}
\bibliography{main.bbl}

\begin{thebibliography}{10}
\providecommand{\url}[1]{#1}
\csname url@samestyle\endcsname
\providecommand{\newblock}{\relax}
\providecommand{\bibinfo}[2]{#2}
\providecommand{\BIBentrySTDinterwordspacing}{\spaceskip=0pt\relax}
\providecommand{\BIBentryALTinterwordstretchfactor}{4}
\providecommand{\BIBentryALTinterwordspacing}{\spaceskip=\fontdimen2\font plus
\BIBentryALTinterwordstretchfactor\fontdimen3\font minus
  \fontdimen4\font\relax}
\providecommand{\BIBforeignlanguage}[2]{{%
\expandafter\ifx\csname l@#1\endcsname\relax
\typeout{** WARNING: IEEEtran.bst: No hyphenation pattern has been}%
\typeout{** loaded for the language `#1'. Using the pattern for}%
\typeout{** the default language instead.}%
\else
\language=\csname l@#1\endcsname
\fi
#2}}
\providecommand{\BIBdecl}{\relax}
\BIBdecl

\bibitem{hsieh2017collaborative}
C.-K. Hsieh, L.~Yang, Y.~Cui, T.-Y. Lin, S.~Belongie, and D.~Estrin,
  ``Collaborative metric learning,'' in \emph{Proc. 26th Int. Conf. {W}orld
  {W}ide {W}eb ({WWW}'17)}, Perth, Aust., Apr. 2017, pp. 193--201.

\bibitem{ebesu2018collaborative}
T.~Ebesu, B.~Shen, and Y.~Fang, ``Collaborative memory network for
  recommendation systems,'' in \emph{Proc. 41st Int. ACM SIGIR Conf. Res.
  Develop. Inf. Retrieval ({SIGIR}'18)}, Ann Arbor, MI, Jul. 2018, pp.
  515--524.

\bibitem{han2019adaptive}
J.~Han, L.~Zheng, Y.~Xu, B.~Zhang, F.~Zhuang, S.~Y. Philip, and W.~Zuo,
  ``Adaptive deep modeling of users and items using side information for
  recommendation,'' \emph{{IEEE} Trans. Neural Netw. Learn. Syst.}, vol.~31,
  no.~3, pp. 737--748, Mar. 2020.

\bibitem{gori2007itemrank}
M.~Gori, A.~Pucci, V.~Roma, and I.~Siena, ``Item{R}ank: {A} random-walk based
  scoring algorithm for recommender engines.'' in \emph{Proc. 20th Int. Joint
  Conf. Artif. Intell. ({IJCAI}'07)}, Hyderabad, India, Jan. 2007, pp.
  2766--2771.

\bibitem{yang2018hop}
J.-H. Yang, C.-M. Chen, C.-J. Wang, and M.-F. Tsai, ``{HOP}-rec: {H}igh-order
  proximity for implicit recommendation,'' in \emph{Proc. 12th ACM Conf.
  Recommender Syst. ({RecSys}'18)}, Vancouver, Canada, Oct. 2018, pp. 140--144.

\bibitem{gao2018bine}
M.~Gao, L.~Chen, X.~He, and A.~Zhou, ``{BiNE: B}ipartite network embedding,''
  in \emph{Proc. 41st Int. ACM SIGIR Conf. Res. Develop. Inf. Retrieval
  ({SIGIR}'18)}, Ann Arbor, MI, Jul. 2018, pp. 715--724.

\bibitem{chen2019collaborative}
C.-M. Chen, C.-J. Wang, M.-F. Tsai, and Y.-H. Yang, ``Collaborative similarity
  embedding for recommender systems,'' in \emph{Proc. 28th Int. Conf. {W}orld
  {W}ide {W}eb ({WWW}'19)}, San Francisco, CA, May 2019, pp. 2637--2643.

\bibitem{kipf2017semi}
T.~N. Kipf and M.~Welling, ``Semi-supervised classification with graph
  convolutional networks,'' in \emph{Proc. 5th Int. Conf. Learn.
  Representations ({ICLR}'17)}, Toulon, France, Apr. 2017.

\bibitem{hamilton2017inductive}
W.~Hamilton, Z.~Ying, and J.~Leskovec, ``Inductive representation learning on
  large graphs,'' in \emph{Proc. 28th Int. Conf. Neural Inf. Process. Syst.
  ({NIPS}’17)}, Long Beach, CA, Dec. 2017, pp. 1025--1035.

\bibitem{xu2018powerful}
K.~Xu, W.~Hu, J.~Leskovec, and S.~Jegelka, ``How powerful are graph neural
  networks?'' in \emph{Proc. 7th Int. Conf. Learn. Representations. (ICLR'19)},
  New Orleans, LA, May 2019.

\bibitem{velivckovic2017graph}
P.~Velickovic, G.~Cucurull, A.~Casanova, A.~Romero, P.~Li{\`{o}}, and
  Y.~Bengio, ``Graph attention networks,'' in \emph{Proc. 6th Int. Conf. Learn.
  Representations. (ICLR'18)}, Vancouver, Canada, Apr.--May 2018.

\bibitem{wu2019survey}
Z.~{Wu}, S.~{Pan}, F.~{Chen}, G.~{Long}, C.~{Zhang}, and P.~S. {Yu}, ``A
  comprehensive survey on graph neural networks,'' \emph{{IEEE} Trans. Neural
  Netw. Learn. Syst.}, vol.~32, no.~1, pp. 4--24, Jan. 2021.

\bibitem{ying2018graph}
R.~Ying, R.~He, K.~Chen, P.~Eksombatchai, W.~L. Hamilton, and J.~Leskovec,
  ``Graph convolutional neural networks for web-scale recommender systems,'' in
  \emph{Proc. 24th {ACM} {SIGKDD} Int. Conf. Knowl. Discovery and Data Mining
  ({KDD}'18)}, London, UK, Aug. 2018, pp. 974--983.

\bibitem{zheng2018spectral}
L.~Zheng, C.-T. Lu, F.~Jiang, J.~Zhang, and P.~S. Yu, ``Spectral collaborative
  filtering,'' in \emph{Proc. 12th ACM Conf. Recommender Syst. ({RecSys}'18)},
  Vancouver, Canada, Oct. 2018, pp. 311--319.

\bibitem{wang2019neural}
X.~Wang, X.~He, M.~Wang, F.~Feng, and T.-S. Chua, ``Neural graph collaborative
  filtering,'' in \emph{Proc. 42nd Int. ACM SIGIR Conf. Res. Develop. Inf.
  Retrieval ({SIGIR}'19)}, Paris, France, Jul. 2019, pp. 165--174.

\bibitem{wang2020multi}
X.~Wang, R.~Wang, C.~Shi, G.~Song, and Q.~Li, ``Multi-component graph
  convolutional collaborative filtering,'' in \emph{Proc. 34th AAAI Conf.
  Artif. Intell. ({AAAI}'20)}, New York, NY, Apr. 2020, pp. 6267--6274.

\bibitem{wu2020joint}
L.~Wu, Y.~Yang, K.~Zhang, R.~Hong, Y.~Fu, and M.~Wang, ``Joint item
  recommendation and attribute inference: An adaptive graph convolutional
  network approach,'' in \emph{Proc. 43rd Int. ACM SIGIR Conf. Res. Develop.
  Inf. Retrieval ({SIGIR}'20)}, Virtual Event, China, Jul. 2020, pp. 679--688.

\bibitem{chen2020revisiting}
L.~Chen, L.~Wu, R.~Hong, K.~Zhang, and M.~Wang, ``Revisiting graph based
  collaborative filtering: A linear residual graph convolutional network
  approach,'' in \emph{Proc. 34th AAAI Conf. Artif. Intell. ({AAAI}'20)}, New
  York, NY, Feb. 2020, pp. 27--34.

\bibitem{he2020lightgcn}
X.~He, K.~Deng, X.~Wang, Y.~Li, Y.~Zhang, and M.~Wang, ``Light{GCN:
  S}implifying and powering graph convolution network for recommendation,'' in
  \emph{Proc. 43rd Int. ACM SIGIR Conf. Res. Develop. Inf. Retrieval
  ({SIGIR}'20)}, Virtual Event, China, Jul. 2020, pp. 639--648.

\bibitem{zhu2020beyond}
J.~Zhu, Y.~Yan, L.~Zhao, M.~Heimann, L.~Akoglu, and D.~Koutra, ``Beyond
  homophily in graph neural networks: Current limitations and effective
  designs,'' in \emph{Proc. 34th Int. Conf. Neural Inf. Process. Syst.
  ({NIPS}’20)}, Virtual Event, Dec. 2020, pp. 7793--7804.

\bibitem{pei2020geom}
H.~Pei, B.~Wei, K.~C.-C. Chang, Y.~Lei, and B.~Yang, ``Geom-{GCN}: {G}eometric
  graph convolutional networks,'' in \emph{Proc. 8th Int. Conf. Learn.
  Representations ({ICLR}'20)}, Virtual Event, Apr. 2020.

\bibitem{zhu2020graph}
J.~Zhu, R.~A. Rossi, A.~Rao, T.~Mai, N.~Lipka, N.~K. Ahmed, and D.~Koutra,
  ``Graph neural networks with heterophily,'' in \emph{Proc. 35th AAAI Conf.
  Artif. Intell. ({AAAI}'21)}, Virtual Event, Feb. 2021, pp. 11\,168--11\,176.

\bibitem{paudel2019loss}
P.~Bibek, L.~Sandro, and B.~Abraham, ``Loss aversion in recommender systems:
  Utilizing negative user preference to improve recommendation quality,'' in
  \emph{Proc. 1st Int. workshop on Context-Aware Recommendation Syst. with Big
  Data Analytics ({CARS-BDA}'19)}, Melbourne, Aust., Feb 2019.

\bibitem{fiftyshades2016}
E.~Frolov and I.~Oseledets, ``Fifty shades of ratings: How to benefit from a
  negative feedback in top-{N} recommendations tasks,'' in \emph{Proc. 12th ACM
  Conf. Recommender Syst. ({RecSys}'16)}, New York, NY, 2016, pp. 91--98.

\bibitem{fewer17}
B.~Paudel, T.~Haas, and A.~Bernstein, ``Fewer flops at the top: Accuracy,
  diversity, and regularization in two-class collaborative filtering,'' in
  \emph{Proc. 12th ACM Conf. Recommender Syst. ({RecSys}'17)}, New York, NY,
  USA, 2017, pp. 215--223.

\bibitem{Tran2021}
K.~H. Tran, A.~Ghazimatin, and R.~Saha~Roy, ``Counterfactual explanations for
  neural recommenders,'' in \emph{Proc. 43rd Int. ACM SIGIR Conf. Res. Develop.
  Inf. Retrieval ({SIGIR}'21)}, Virtual Event, Jul. 2021, pp. 1627--1631.

\bibitem{derr2018signed}
T.~Derr, Y.~Ma, and J.~Tang, ``Signed graph convolutional networks,'' in
  \emph{Proc. 18th IEEE Int. Conf. Data Mining ({ICDM}'18)}, Singap., Nov.
  2018, pp. 929--934.

\bibitem{huang2019signed}
J.~Huang, H.~Shen, L.~Hou, and X.~Cheng, ``Signed graph attention networks,''
  in \emph{Proc. 28th Int. Conf. Artif. Neural Netw. ({ICANN}'19)}, Munich,
  Germany, Sep. 2019, pp. 566--577.

\bibitem{heider1946attitudes}
F.~Heider, ``Attitudes and cognitive organization,'' \emph{J. psychol.},
  vol.~21, no.~1, pp. 107--112, 1946.

\bibitem{BPRMF09}
S.~Rendle, C.~Freudenthaler, Z.~Gantner, and L.~Schmidt-Thieme, ``{BPR:
  B}ayesian personalized ranking from implicit feedback,'' in \emph{Proc. 25th
  Conf. Uncertainty Artif. Intell. ({UAI}'09)}, Montreal, Canada, Jun. 2009,
  pp. 452--–461.

\bibitem{koren2008factorization}
Y.~Koren, ``Factorization meets the neighborhood: {A} multifaceted
  collaborative filtering model,'' in \emph{Proc. 14th ACM SIGKDD Int. Conf.
  Knowl. Discovery and Data Mining ({KDD}'08)}, Las Vegas, NV, Aug. 2008, pp.
  426--434.

\bibitem{koren2009matrix}
Y.~Koren, R.~Bell, and C.~Volinsky, ``Matrix factorization techniques for
  recommender systems,'' \emph{IEEE Computer}, vol.~42, no.~8, pp. 30--37,
  2009.

\bibitem{luo2015nonnegative}
X.~Luo, M.~Zhou, S.~Li, Z.~You, Y.~Xia, and Q.~Zhu, ``A nonnegative latent
  factor model for large-scale sparse matrices in recommender systems via
  alternating direction method,'' \emph{{IEEE} Trans. Neural Netw. Learn.
  Syst.}, vol.~27, no.~3, pp. 579--592, 2016.

\bibitem{lin2007projected}
C.-J. Lin, ``Projected gradient methods for nonnegative matrix factorization,''
  \emph{Neural Comput.}, vol.~19, no.~10, pp. 2756--2779, 2007.

\bibitem{xue2017deep}
H.~Xue, X.~Dai, J.~Zhang, S.~Huang, and J.~Chen, ``Deep matrix factorization
  models for recommender systems.'' in \emph{Proc. 26th Int. Joint Conf. Artif.
  Intell. ({IJCAI}'17)}, Melbourne, Aust., Aug. 2017, pp. 3203--3209.

\bibitem{he2017neural}
X.~He, L.~Liao, H.~Zhang, L.~Nie, X.~Hu, and T.~Chua, ``Neural collaborative
  filtering,'' in \emph{Proc. 26th Int. Conf. {W}orld {W}ide {W}eb ({WWW}'17)},
  Perth, Aust., Apr. 2017, pp. 173--182.

\bibitem{ning2011slim}
X.~Ning and G.~Karypis, ``{SLIM: S}parse linear methods for top-{N} recommender
  systems,'' in \emph{Proc. 11th {IEEE} Int. Conf. Data Mining ({ICDM}'11)},
  Vancouver, Canada, Dec. 2011, pp. 497--506.

\bibitem{kabbur2013fism}
S.~Kabbur, X.~Ning, and G.~Karypis, ``{FISM: F}actored item similarity models
  for top-{N} recommender systems,'' in \emph{Proc. 19th {ACM} {SIGKDD} Int.
  Conf. Knowl. Discovery and Data Mining ({KDD}'13)}.\hskip 1em plus 0.5em
  minus 0.4em\relax Chicago, IL: {ACM}, Aug. 2013, p. 659–667.

\bibitem{nguyen2018npe}
T.~Nguyen and A.~Takasu, ``{NPE: N}eural personalized embedding for
  collaborative filtering,'' in \emph{Proc. 27th Int. Joint Conf. Artif.
  Intell. ({IJCAI}'18)}, Stockholm, Sweden, Jul. 2018, pp. 1583--1589.

\bibitem{he2018nais}
X.~He, Z.~He, J.~Song, Z.~Liu, Y.-G. Jiang, and T.-S. Chua, ``{NAIS: N}eural
  attentive item similarity model for recommendation,'' \emph{IEEE Trans.
  Knowl. Data Eng.}, vol.~30, no.~12, pp. 2354--2366, Dec. 2018.

\bibitem{deng2016deep}
S.~Deng, L.~Huang, G.~Xu, X.~Wu, and Z.~Wu, ``On deep learning for trust-aware
  recommendations in social networks,'' \emph{{IEEE} Trans. Neural Netw. Learn.
  Syst.}, vol.~28, no.~5, pp. 1164--1177, 2017.

\bibitem{van2018graph}
R.~van~den Berg, T.~N. Kipf, and M.~Welling, ``Graph convolutional matrix
  completion,'' in \emph{Proc. KDD'18 Deep Learning Day}, 2018.

\bibitem{zhang2019inductive}
M.~Zhang and Y.~Chen, ``Inductive matrix completion based on graph neural
  networks,'' in \emph{Proc. 8th Int. Conf. Learn. Representations
  ({ICLR}'20)}, Virtual Event, Apr. 2020.

\bibitem{wu2019neural}
L.~Wu, P.~Sun, Y.~Fu, R.~Hong, X.~Wang, and M.~Wang, ``A neural influence
  diffusion model for social recommendation,'' in \emph{Proc. 42nd Int. ACM
  SIGIR Conf. Res. Develop. Inf. Retrieval ({SIGIR}'19)}, Paris, France, Jul.
  2019, pp. 235--244.

\bibitem{fan2019graph}
W.~Fan, Y.~Ma, Q.~Li, Y.~He, E.~Zhao, J.~Tang, and D.~Yin, ``Graph neural
  networks for social recommendation,'' in \emph{Proc. 28th Int. Conf. {W}orld
  {W}ide {W}eb ({WWW}'19)}, San Francisco, CA, May 2019, pp. 417--426.

\bibitem{wu2020diffnet++}
L.~Wu, J.~Li, P.~Sun, R.~Hong, Y.~Ge, and M.~Wang, ``{DiffNet++: A} neural
  influence and interest diffusion network for social recommendation,''
  \emph{{IEEE} Trans. Knowl. and Data Eng.}, to appear.

\bibitem{chen2020neural_HGCN}
X.~Chen, K.~Xiong, Y.~Zhang, L.~Xia, D.~Yin, and J.~X. Huang, ``Neural
  feature-aware recommendation with signed hypergraph convolutional network,''
  \emph{ACM Trans. Inf. Syst.}, vol.~39, no.~1, pp. 1--22, Nov. 2020.

\bibitem{wang2017signed}
S.~Wang, J.~Tang, C.~Aggarwal, Y.~Chang, and H.~Liu, ``Signed network embedding
  in social media,'' in \emph{Proc. SIAM Int. Conf. Data Mining (SIAM'17)},
  Houston, Texas, Apr. 2017, pp. 327--335.

\bibitem{kim2018side}
J.~Kim, H.~Park, J.-E. Lee, and U.~Kang, ``{SIDE: R}epresentation learning in
  signed directed networks,'' in \emph{Proc. 27th Int. Conf. {W}orld {W}ide
  {W}eb ({WWW}'18)}, Lyon, France, Apr. 2018, pp. 509--518.

\bibitem{li2020learning}
Y.~Li, Y.~Tian, J.~Zhang, and Y.~Chang, ``Learning signed network embedding via
  graph attention,'' in \emph{Proc. 34th AAAI Conf. Artif. Intell.
  ({AAAI}'20)}, vol.~34, no.~04, New York, NY, USA, Apr. 2020, pp. 4772--4779.

\bibitem{liu2021signed}
H.~Liu, Z.~Zhang, P.~Cui, Y.~Zhang, Q.~Cui, J.~Liu, and W.~Zhu, ``Signed graph
  neural network with latent groups,'' in \emph{Proc. 27th ACM SIGKDD Int.
  Conf. Knowl. Discovery and Data Mining ({KDD}'21)}, Virtual Event, Singap.,
  Aug. 2021, pp. 1066--1075.

\bibitem{leskovec2010predicting}
J.~Leskovec, D.~Huttenlocher, and J.~Kleinberg, ``Predicting positive and
  negative links in online social networks,'' in \emph{Proc. 19th Int. Conf.
  {W}orld {W}ide {W}eb ({WWW}'10)}, Raleigh, NC, Apr. 2010, pp. 641--650.

\bibitem{yuan2017sne}
S.~Yuan, X.~Wu, and Y.~Xiang, ``{SNE: S}igned network embedding,'' in
  \emph{Proc. 21st Pacific-Asia Conf. Advances Knowl. and Data Mining
  ({PAKDD}'17)}, Jeju, South Korea, May 2017, pp. 183--195.

\bibitem{jin2021node}
W.~Jin, T.~Derr, Y.~Wang, Y.~Ma, Z.~Liu, and J.~Tang, ``Node similarity
  preserving graph convolutional networks,'' in \emph{Proc. 14th ACM Int. Conf.
  Web Search and Data Mining ({WSDM'21})}, Virtual Event, Isr., Mar. 2021, pp.
  148--156.

\bibitem{vaswani2017attention}
A.~Vaswani, N.~Shazeer, N.~Parmar, J.~Uszkoreit, L.~Jones, A.~N. Gomez,
  {\L}.~Kaiser, and I.~Polosukhin, ``Attention is all you need,'' in
  \emph{Proc. 31st Int. Conf. Neural Inf. Process. Syst. ({NIPS}’17)}, Long
  Beach, CA, Dec. 2017, pp. 5998--6008.

\bibitem{gilmer2017neural}
J.~Gilmer, S.~S. Schoenholz, P.~F. Riley, O.~Vinyals, and G.~E. Dahl, ``Neural
  message passing for quantum chemistry,'' in \emph{Proc. 34th Int. Conf. Mach.
  Learn. (ICML'17)}, Sydney, Aust., Aug. 2017, pp. 1263--1272.

\bibitem{xu2018representation}
K.~Xu, C.~Li, Y.~Tian, T.~Sonobe, K.~Kawarabayashi, and S.~Jegelka,
  ``Representation learning on graphs with jumping knowledge networks,'' in
  \emph{Proc. 35th Int. Conf. Mach. Learn. (ICML'18)}, Stockholm, Sweden, Jul.
  2018, pp. 5449--5458.

\bibitem{li2018deeper}
Q.~Li, Z.~Han, and X.-M. Wu, ``Deeper insights into graph convolutional
  networks for semi-supervised learning,'' in \emph{Proc. 32nd AAAI Conf.
  Artif. Intell. ({AAAI}'18)}, New Orleans, LA, Feb. 2018, pp. 3538--3545.

\bibitem{mikolovskipgram}
T.~Mikolov, I.~Sutskever, K.~Chen, G.~S. Corrado, and J.~Dean, ``Distributed
  representations of words and phrases and their compositionality,'' in
  \emph{Proc. 27th Conf. Neural Inf. Process. Systems (NIPS'13)}, Lake Tahoe,
  NV, Dec. 2013, pp. 3111--3119.

\bibitem{tang2015line}
J.~Tang, M.~Qu, M.~Wang, M.~Zhang, J.~Yan, and Q.~Mei, ``{LINE: L}arge-scale
  information network embedding,'' in \emph{Proc. 24th Int. Conf. {W}orld
  {W}ide {W}eb (WWW'15)}, Florence, Italy, May 2015, pp. 1067--1077.

\bibitem{harper2015movielens}
F.~M. Harper and J.~A. Konstan, ``The movielens datasets: {H}istory and
  context,'' \emph{{ACM} Trans. Interactive Intell. Syst.}, vol.~5, no.~4, pp.
  1--19, Dec. 2015.

\bibitem{he2016ups}
R.~He and J.~McAuley, ``Ups and downs: Modeling the visual evolution of fashion
  trends with one-class collaborative filtering,'' in \emph{Proc. 25th Int.
  Conf. {W}orld {W}ide {W}eb ({WWW}'16)}, Montreal, Canada, Apr. 2016, pp.
  507--517.

\bibitem{he2016vbpr}
------, ``{VBPR: V}isual bayesian personalized ranking from implicit
  feedback,'' in \emph{Proc. 30th AAAI Conf. Artif. Intell. ({AAAI}'16)},
  Phoenix, AZ, Feb. 2016, pp. 144--150.

\bibitem{NDCG}
K.~J\"{a}rvelin and J.~Kek\"{a}l\"{a}inen, ``Cumulated gain-based evaluation of
  {IR} techniques,'' \emph{ACM Trans. Inf. Syst.}, vol.~20, no.~4, pp.
  422--446, Oct. 2002.

\bibitem{fey2019pyg}
M.~Fey and J.~E. Lenssen, ``Fast graph representation learning with pytorch
  geometric,'' in \emph{Proc. ICLR Workshop on Representation Learning on
  Graphs and Manifolds}, New Orleans, LA, May 2019.

\bibitem{xavier_}
X.~Glorot and Y.~Bengio, ``Understanding the difficulty of training deep
  feedforward neural networks,'' in \emph{Proc. 13th Int. Conf. Artif. Intell.
  Statist. ({AISTATS}'10)}, Sardinia, Italy, May 2010, pp. 249--256.

\bibitem{srivastava2014dropout}
N.~Srivastava, G.~Hinton, A.~Krizhevsky, I.~Sutskever, and R.~Salakhutdinov,
  ``Dropout: {A} simple way to prevent neural networks from overfitting,''
  \emph{J. Mach. Learn. Res.}, vol.~15, no.~1, pp. 1929--1958, Jan. 2014.

\bibitem{kingma2015adam}
D.~P. Kingma and J.~Ba, ``Adam: {A} method for stochastic optimization,'' in
  \emph{Proc. 3rd Int. Conf. Learn. Representations ({ICLR}'15)}, San Diego,
  CA, May 2015.

\end{thebibliography}
\end{document}